\documentclass[useAMS,usenatbib]{mn2e}
\usepackage{psfig}
\usepackage{amssymb}
\usepackage{color}
\usepackage{enumerate}
\usepackage{rotating}
\usepackage{amsmath}
\usepackage{mathrsfs}
\usepackage{ragged2e}
\usepackage{hyperref}
\usepackage{comment}
\usepackage{float}
\usepackage{relsize}
\newcommand{\be}{\begin{equation}}
\newcommand{\ee}{\end{equation}}
\newcommand{\bea}{\begin{eqnarray}}
\newcommand{\eea}{\end{eqnarray}}
\newcommand{\no}{\nonumber}

\title[On the kSZ effect as a probe for spin bias]
{On the kinetic Sunyaev-Zel'dovich effect as an observational probe for halo spin bias \\
}

\author[Montero-Dorta et al.]{
\parbox[t]{\textwidth}{
Antonio D. Montero-Dorta$^{1}$\thanks{E-mail: amonterodorta@gmail.com}, M. Celeste Artale$^{2}$, L. Raul Abramo$^{1}$, Beatriz Tucci$^{1}$} 
\vspace*{6pt} \\ 
$^1$ Departamento de F\'isica Matem\'atica, Instituto de F\'isica, Universidade de S\~ao Paulo, Rua do Mat\~ao 1371, CEP 05508-090, \\
S\~ao Paulo, Brazil \\
$^2$ Institut f\"ur Astro- und Teilchenphysik, Universit\"at Innsbruck, Technikerstrasse 25/8, 6020 Innsbruck, Austria \\
\vspace{-0.4cm} 
}

\date{Accepted ---. Received ---;in original form --- \vspace{-0.3cm}}

\def\simlt{\lower.5ex\hbox{$\; \buildrel < \over \sim \;$}}
\def\simgt{\lower.5ex\hbox{$\; \buildrel > \over \sim \;$}}
\usepackage{graphicx}
\usepackage{rotating}

\definecolor{red}{rgb}{1,0,0}

\begin{document}

\bibliographystyle{mnras}

\maketitle

\begin{abstract}
We explore the potential of the {\it{kinetic}} Sunyaev-Zel'dovich (kSZ) effect as the cornerstone of a future observational probe for {\it{halo spin bias}}, the secondary dependence of halo clustering on halo spin at fixed halo mass. Using the IllustrisTNG magneto-hydrodynamical cosmological simulation, we measure both the kSZ and the {\it{thermal}} SZ (tSZ) effects produced by the baryonic content of more than 50,000 haloes within the halo mass range $11 < \log_{10} ({\rm M_{vir}}/ h^{-1} {\rm M_{\odot}}) \lesssim 14.5$. First, we confirm that the magnitude of both effects depends strongly on the total gas and virial mass of the haloes, and that the integrated kSZ signal displays a significant correlation with the angular momentum of the intra-halo gas, particularly for massive haloes. Second, we show that both the integrated kSZ signal and the ratio of the integrated kSZ and tSZ signals trace total halo spin, even though significant scatter exists. Finally, we demonstrate that, in the absence of observational and instrumental uncertainties, these SZ-related statistics can be used to recover most of the underlying IllustrisTNG halo spin bias signal. Our analysis represents the first attempt to develop a future observational probe for halo spin bias, bringing forward alternative routes for measuring the secondary bias effects.

\end{abstract}


\section{Introduction}
\label{sec:intro}

Among the well-established secondary dependencies of halo clustering, the one that has drawn more attention is the dependence on halo age, or, more generally, the assembly history of haloes, an effect called {\it{halo assembly bias}}. At fixed halo mass, lower-mass dark-matter (DM) haloes that assemble a significant fraction of their mass early on are more tightly clustered than those that assemble it at later times, with the signal progressively vanishing towards the high-mass end. Halo assembly bias is, however, just a particular case of the more general {\it{secondary halo bias}}, which includes dependencies on multiple halo properties (see, e.g., \citealt{Sheth2004,gao2005,wechsler2006,Gao2007,Angulo2008,li2008,Lazeyras2017,salcedo2018,han2018,Mao2018, SatoPolito2019, Johnson2019, MonteroDorta2020B, Tucci2020}). Among these other important dependencies, the secondary dependence on halo spin, called {\it{halo spin bias}}, is the focus of this work.

Although halo spin bias was first observed more than a decade ago (e.g., \citealt{Gao2007}), only recent measurements have provided a complete description of the signal. \cite{SatoPolito2019}, in particular, presented an accurate measurement of spin bias for a halo mass range spanning four orders of magnitude. While it was known from early on that higher-spin haloes are more tightly clustered than lower-spin haloes at the high-mass end, \cite{SatoPolito2019} showed that the trend actually inverts below a characteristic mass. At $z=0$, slower rotators are in fact more highly biased than faster rotators of the same mass below $\log_{10} ({\rm M_{vir}}/h^{-1} {\rm M_{\odot}}) \sim 11.5$, with this characteristic mass shifting towards smaller masses at higher redshifts \citep{Tucci2020}. This {\it{spin bias inversion}}, which was subsequently confirmed by \citealt{Johnson2019}, is caused by the 
effect of {\it{splashback}} haloes, as shown in \cite{Tucci2020}. The splashback haloes are distinct haloes at the redshift under analysis that previously passed through the virial radius of other distinct but more massive haloes, thus sharing the large-scale bias properties of their companions. Although the physical origin of the spin bias inversion has been revealed, a compelling theory for the {\it{intrinsic}} mass dependence of the signal, once splashbacks are removed, is yet to be established (see dicussion in, e.g., \citealt{Lacerna2012,salcedo2018,Paranjape2018, Johnson2019, Ramakrishnan2019, SatoPolito2019, Mansfield2020, Tucci2020}).

The above halo bias dependencies are particularly interesting if they produce a measurable imprint on the galaxy population. As shown in \cite{MonteroDorta2020B} using the IllustrisTNG magneto-hydrodynamical simulation\footnote{\url{http://www.tng-project.org}}, these effects are expected to be transmitted to (or traced by) the central galaxy population, in the sense that the clustering of central galaxies {\it{at fixed halo mass}} should depend on secondary halo properties such as halo formation time. The existence of this so-called {\it{galaxy assembly bias}}\footnote{Note that, in this context, it is common to refer to all the secondary dependencies of galaxy clustering as {\it{galaxy assembly bias}}, although, strictly speaking, the name {\it{secondary galaxy bias}} would be more adequate.} has important consequences, not only in terms of the halo--galaxy connection, but also for structure formation, the modelling of halo clustering, and the extraction of cosmological information from galaxy surveys (see discussion in  \citealt{Hearin2013,Hearin2014,Hearin2016,Zentner2019,Wechsler2018}). Although several attempts have been reported, conclusive evidence for the existence of galaxy assembly bias has not been established yet \citep{Zentner2016,Miyatake2016,Zu2016, Lin2016,Sunayama2016,MonteroDorta2017B,Niemiec2018,Walsh2019,Sunayama2019,Obuljen2020}.    

In the context of a potential observational detection, halo spin bias has the advantage that the signal itself is expected to increase with halo mass (see, e.g., \citealt{SatoPolito2019,Johnson2019,Tucci2020}). The potential signal should, therefore, be maximal for the biggest clusters, which are in turn easier to probe. Probing halo spin is, however, intrinsically challenging, as it represents a second-order measurement with respect to halo mass. In this work, we use the IllustrisTNG simulation at redshift $z=0$ to investigate the observability of spin bias through the rotation of the intra-halo gas (or intra-cluster medium, ICM\footnote{Note that, strictly speaking, the haloes that we analyse in this work span a wide range of halo masses that is not restricted to cluster-size haloes. For this reason, the term ``intra-halo gas" is favoured throughout this work.}), which is expected to trace the rotation of the DM component of the halo (e.g. \citealt{MonteroDorta2020B}). These measurements are combined with a clustering analysis performed at fixed halo mass, a type of study that has only become feasible in hydrodynamical simulations recently, thanks to the increase in cosmological volume provided by some of the IllustrisTNG boxes.

The intra-halo gas is defined in this work as all the gas that is contained inside the halo, including the amount that is in the form of galaxies. In order to probe its global rotation, we measure the Sunyaev Zel'dovich effect (SZ, \citealt{SZ1970,SZ1980A,SZ1980B}) induced by every single halo with mass M$_{\rm vir} > 10^{11}$ $h^{-1}$M$_{\odot}$ in the IllustrisTNG100 and IllustrisTNG300 boxes (hereafter TNG100 and TNG300, respectively). The SZ effect is produced by the inverse Compton scattering of the photons of the cosmic microwave background (CMB) as they propagate through galaxy clusters. The  observational probe that we explore here is based on the {\it{kinetic}} SZ (kSZ) effect, where the scattering is due to the coherent, bulk motion of electrons inside the cluster. A purely rotational motion of the ICM would produce variations in the CMB spectrum (or in temperature, i.e., $\Delta T/T$) in the form of a dipole (see \citealt{Cooray2002, Chluba2002}). Quantitatively, the kSZ signal is proportional to the integral of the product of the electron number density and the cluster velocity component along the line of sight ({\it{los}}), so it is expected to be higher from richer (i.e., more massive) and faster-rotating haloes. We complement our kSZ measurement with the {\it{thermal}} SZ (tSZ) effect, in which the Compton scattering of CMB photons is produced by the random thermal motion of electrons. The tSZ effect is a well-documented proxy for the total mass of the halo \citep[see e.g.,][]{Arnaud2010,Andersson2011,Marriage2011,Sifon2013,Battaglia2012,Krause2012,Sembolini2013,Yu2015,Lim2020}, which is of course a critical quantity for secondary bias studies.

Analytical models of the kSZ effect (\citealt{Chluba2002,Cooray2002}) predict amplitudes at the peak of the dipole of the order of a few to a few dozen $\mu$K, this depending, of course, on the orientation of the halo's rotational axis with respect to the {\it{los}}, and on the properties of the cluster itself. Observationally, measuring the kSZ effect from individual objects is still challenging (e.g., \citealt{Dupke2002,Mroczkowski2012, Sayers2013, Adam2017,Sayers2019}), but significant progress has been made in recent years using a more {\it{statistical}} approach. In particular, the kSZ effect has been detected at high statistical significance using the relative pairwise momentum between the kSZ signals produced by pairs of galaxy clusters (see, e.g., \citealt{Hand2012,Planck_kSZ2016,Li_kSZ2018}). Next-generation instrumentation, however, is expected to allow measurements on an object-by-object basis for sizeable samples of clusters, which will have important applications for cosmology (e.g., \citealt{HM2006,Bhattacharya2007,Bhattacharya2008,Zhang2011,Hand2012, Eva2015, Alonso2016, Soergel2018, Basu2019, Mroczkowski2019, Andrina2020}).

Hydrodynamical simulations are laboratories for galaxy-formation physics and excellent forecasting tools for future observational probes. \cite{Baldi2018} demonstrated, using a small sample of haloes from the MUSIC (Marenostrum-MultiDark SImulations of galaxy Clusters) high-resolution simulation \citep{Sembolini2013}, not only that the amplitude of the kSZ temperature distortion in hydrodynamical simulations is consistent with theoretical estimates, but also, that the expected radial profile of the rotational velocity of the gas can be reconstructed from the measured signal. More recently, \cite{Lim2020} provided a detailed comparison of the physical properties of the intra-halo gas between observations and several hydrodynamical simulations. These works, and other related halo--galaxy connection analyses using hydrodynamical simulations, makes us confident about the validity of our forecast. 

Following the philosophy of \cite{Baldi2018}, we approach the SZ measurements from a ``semi-theoretical" perspective. On the one hand, we use a state-of-the-art hydrodynamical simulation that models realistically the structure and dynamics of gas inside haloes. On the other hand, we neglect, for the time being, the observational and instrument-related uncertainties associated with a real-life detection. The main goal of this first analysis is to discuss and evaluate the potentialities of the SZ effect as an observational probe for halo spin bias, by measuring the {\it{intrinsic signal}} for a large number of hydrodynamical haloes. 

The paper is organised as follows. Section~\ref{sec:sims} provides a brief description of the simulation box and the halo sample analysed in this work. The halo spin bias measurement is reviewed in Section~\ref{sec:spin_bias}. The observational probes, i.e., the SZ effects, are explained in detail in Section~\ref{sec:kSZ}: theory (\ref{sec:theory}), implementation in IllustrisTNG (\ref{sec:method}), and integrated signal (statistic) definition (\ref{sec:signal}). The main results of our analysis are presented in Section~\ref{sec:results}, including: kSZ maps (\ref{sec:maps}), scaling relations (\ref{sec:scaling}), and the clustering analysis (\ref{sec:clustering}). For reference, the current state of SZ observations is summarised in Section~\ref{sec:discussion}. Finally, Section~\ref{sec:conclusions} is devoted to discussing the implications of our results and providing a brief summary of the paper. The IllustrisTNG simulations adopt the standard $\Lambda$CDM cosmology  \citep{Planck2016}, with parameters $\Omega_{\rm m} = 0.3089$,  $\Omega_{\rm b} = 0.0486$, $\Omega_\Lambda = 0.6911$, $H_0 = 100~\,h\, {\rm km\, s^{-1}Mpc^{-1}}$ with $h=0.6774$, $\sigma_8 = 0.8159$ and $n_s = 0.9667$.

\section{Simulation data and halo sample}
\label{sec:sims}

We use the galaxy and DM-halo catalogues from the IllustrisTNG suite of  magneto-hydrodynamical cosmological simulations \citep{Springel2018,Pillepich2018}.
The IllustrisTNG simulations are produced with the {\sc arepo} moving-mesh code \citep{Springel2010} and are built from the Illustris simulation, improving upon its performance \citep{Vogelsberger2014a,Vogelsberger2014b}. The simulation includes sub-grid models for star formation, stellar and AGN feedback, radiative metal-line gas cooling, and chemical enrichment from SNII, SNIa, and AGB stars. It adopts, as mentioned before, a cosmology consistent with the 2015 Planck constraints  \citep{Planck2016}.

In this work, we analyse two different boxes from the IllustrisTNG suite: IllustrisTNG300-1 (that we simply call TNG300), which corresponds to a periodic cubic box of side length 205 $h^{-1}$Mpc, and IllustrisTNG100-1 (TNG100), of  side length 75 $h^{-1}$Mpc. The TNG300 and TNG100 boxes were run from a starting redshift $z = 127$ to the present, $z = 0$. TNG300 uses DM particles of mass $4.0 \times 10^7$ $h^{-1} {\rm M_{\odot}}$, while the mass of the initial gas cells is $7.6 \times 10^6 ~h^{-1} {\rm M_{\odot}}$. TNG100 has higher resolution, with DM particles of mass $5.1 \times 10^6 ~h^{-1} {\rm M_{\odot}}$ and initial gas cells of $9.4 \times 10^5 ~h^{-1} {\rm M_{\odot}}$. The main results of this paper are obtained with TNG300, which provides enough cosmological volume to ensure a reliable clustering measurement. TNG100, on the other hand, allows us to check the robustness of our results against potential resolution issues (see the Appendix). 

\begin{figure}
\includegraphics[width=\columnwidth]{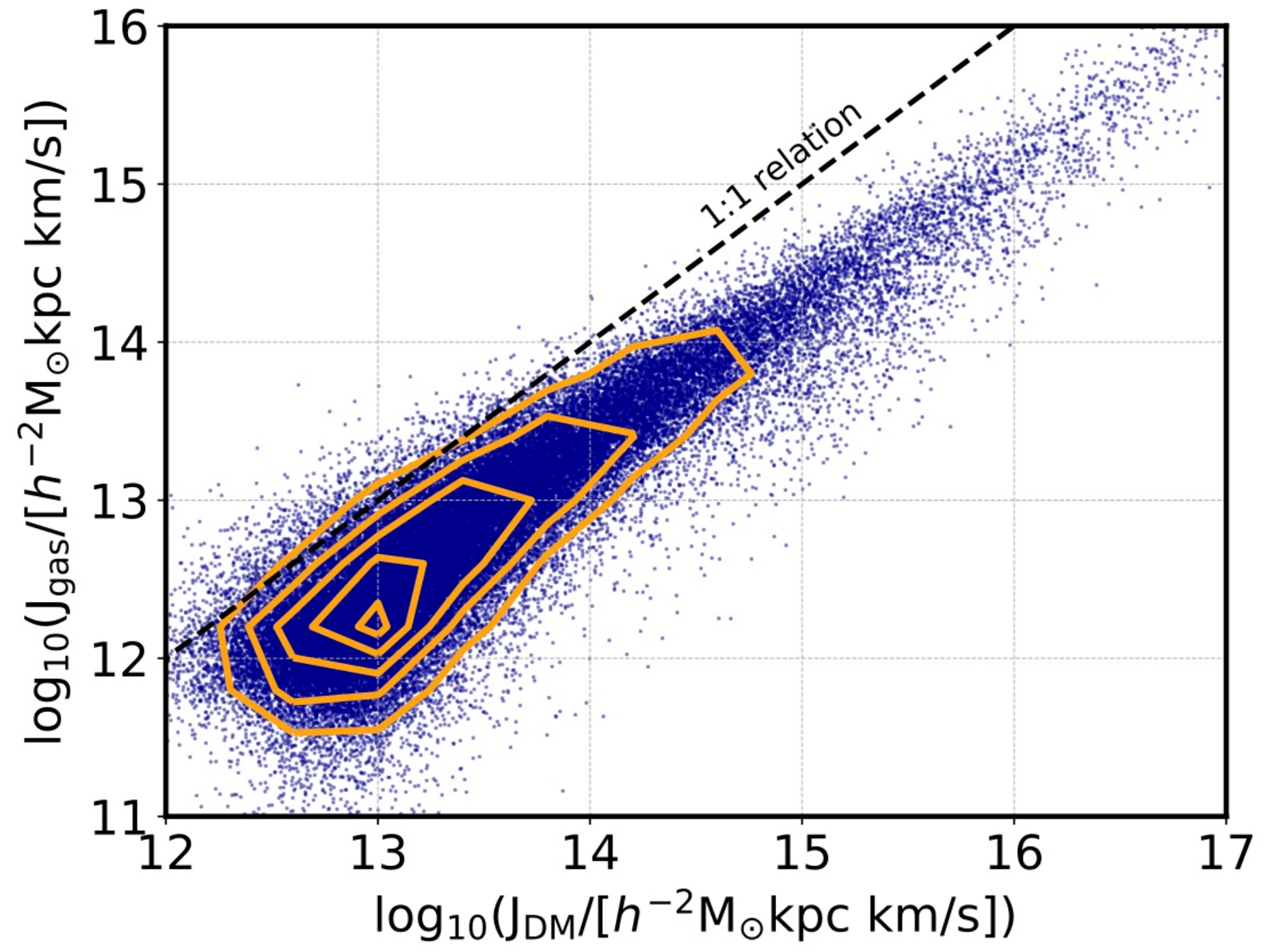}
\caption{The correlation between the angular momentum of the intra-halo gas and that of the DM component within the virial radius, R$_{\rm vir}$, for haloes in TNG300. Contours are computed for all haloes with halo mass M$_{\rm vir} > 10^{10.5}$ $h^{-1}$M$_{\odot}$, whereas dots represent a randomly selected subset containing 10$\%$ of this population. The dashed line indicates the one-to-one relation.}
\label{fig:j_halo}
\end{figure}

In the IllustrisTNG simulations, DM haloes are identified using a friends-of-friends (FOF) algorithm with a linking length of 0.2 times the mean inter-particle separation \citep{Davis1985}. The gravitationally bound substructures (subhaloes) are identified using the {\sc SUBFIND} algorithm \citep{Springel2001,Dolag2009}, with those containing a baryonic component being considered galaxies. Most haloes (or {\it{groups}} in the IllustrisTNG nomenclature) contain multiple galaxies, where the ``central" galaxy is assumed to be the most massive (and gravitationally bound) among all galaxies in the group.

In our analysis, we use the virial mass of the haloes, M$_{\rm vir}$, defined as the {\it{total}} mass enclosed within a sphere of radius R$_{\rm vir}$ (i.e., the radius at which the enclosed density equals 200 times the critical density).
The spin of the halo is defined as in \cite{Bullock2001}, namely:
  \begin{equation}\label{eq:spin}
    \lambda_{\rm halo} = \frac{\rm |J|}{\sqrt{2} {\rm M_{vir}} {\rm V_{vir}} {\rm R_{vir}}},
  \end{equation}
\noindent where J is the angular momentum of the halo and V$_{\rm vir}$ is its circular velocity at the virial radius R$_{\rm vir}$. Note that both M$_{\rm vir}$ and $\lambda_{\rm halo}$ are dominated by the DM component of the halo, even though they are defined for the combination of all halo components (DM+baryons). As shown in \cite{MonteroDorta2020B} (Fig. 3), the inclusion of baryons has little effect on the halo spin bias signal.

We complement our analysis of IllustrisTNG with the MultiDark\footnote{\url{http://skiesanduniverses.org}} suite of N-body cosmological simulations \citep{multidark2016}. Thanks to their size and DM resolution, the MultiDark boxes have produced the most accurate measurements of halo spin bias for a wide halo mass range (see \citealt{SatoPolito2019}). We use 5 different simulation boxes here: Very Small MultiDark Planck (VSMDPL), Small MultiDark Planck (SMDPL), MultiDark Planck 2 (MDPL2), Big MultiDark Planck (BigMDPL) and Huge MultiDark Planck (HugeMDPL). These boxes contain $\sim$4000$^3$ particles and span side lengths of 0.16, 0.4, 1, 2.5 and 4 $h^{-1}$Gpc, respectively. A summary of the numerical parameters of each simulation can be found in \cite{multidark2016}. Haloes in the MultiDark boxes were identified using the {\sc ROCKSTAR} software \citep{Behroozi2013}. Throughout this work, only distinct haloes are considered. 

We finish this section by showing, in Figure~\ref{fig:j_halo}, the strong correlation that exists in TNG300 between the angular momentum of the gas inside haloes and the angular momentum of the the DM component. This plot, which was presented in \cite{MonteroDorta2020B}, motivates the current analysis, since it demonstrates that the dynamics of the DM component are transmitted to the rotating intra-halo gas. Note that, inside the haloes, we make no distinction between the gas that is bound to the galaxies and the gas that distributes around them, since both contribute to the SZ signal.

\section{The halo spin bias measurement}
\label{sec:spin_bias}

\begin{figure}
\includegraphics[width=\columnwidth]{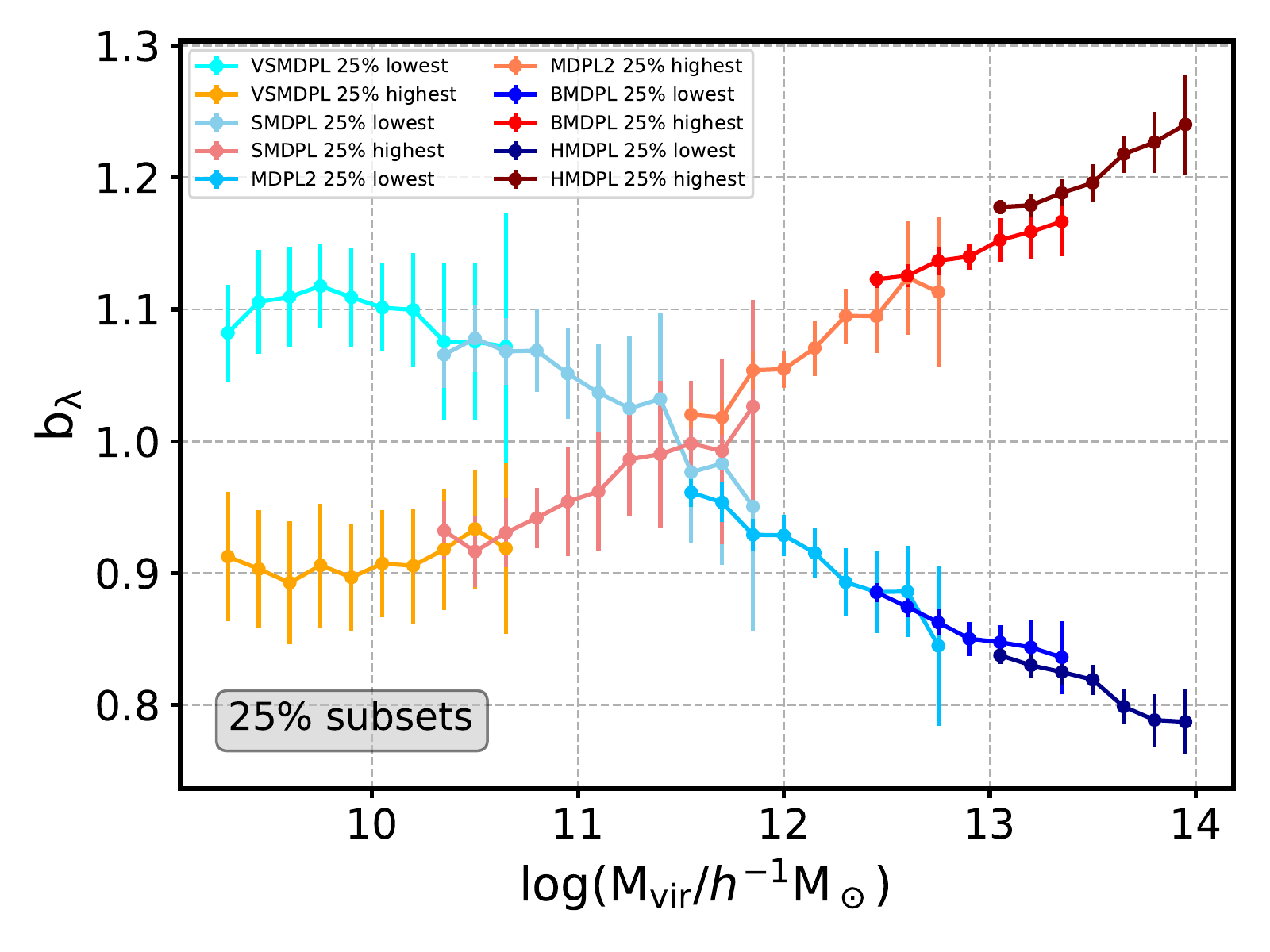}
\caption{The halo spin bias measurement from the MultiDark suite of N-body numerical simulations at $z=0$. Each point represents the relative bias between the high-spin (redder colours) or the low-spin (bluer colours) quartiles and the entire population at a given halo mass bin. Error bars show the standard deviation computed from a set of sub-boxes (see \citealt{SatoPolito2019,Tucci2020} for more details). The MultiDark boxes employed are distinguished by progressively darker tones. From left to right, results from the Very Small MultiDark (VSMD), Small MultiDark (SMD), MultiDark Plack 2 (MDPL2), Big MultiDark (BMD) and Huge MultiDark (HMD) boxes are shown.}
\label{fig:spin_bias}
\end{figure}

Figure~\ref{fig:spin_bias} displays the halo spin bias measurement at $z=0$ from the MultiDark suite of N-body numerical simulations, in the standard way of presenting this secondary dependence (see, e.g., \citealt{SatoPolito2019,Johnson2019,Tucci2020}). The relative spin bias, $b_{\rm \lambda}$, between subsets of haloes selected according to $\lambda_{\rm halo}$ and the entire halo population at a given mass bin is shown as a function of halo mass. These subsets in N-body simulations are usually chosen so that they encompass the 25$\%$ highest- and lowest-spin halo subpopulation, respectively, in order to maximise the signal (redder and bluer tones, respectively, in Figure~\ref{fig:spin_bias}). The
relative bias is measured here using the correlation function. At a given mass bin, M$_i$, and scale r, the value of b$_{\rm \lambda}$ between a $\lambda_{\rm halo}$-subset, $\mathcal{S}_{\rm \lambda}$, and the entire mass bin, can be simply measured as the ratio: 

\begin{equation}
   {\rm b}^2_{\rm \lambda}({\rm r,\mathcal{S}_{\lambda}|M_i}) = \frac{\xi_{[{\rm \mathcal{S}_{\lambda}, \mathcal{S}_{\lambda}}]}({\rm r})}{\xi_{[{\rm M_i,M_i}]}({\rm r})},
   \label{eq:bias}
\end{equation}

\noindent where $\xi$ indicates the auto-correlation, and $S_{\rm \lambda}$ $\in$ M$_i$. In Figure~\ref{fig:spin_bias}, the bias is averaged on scales 5--15 $h^{-1}$Mpc, which is the standard range of scales used in secondary bias analyses. The error bars are obtained as the standard deviation of a set of sub-boxes \citep{SatoPolito2019,Tucci2020}. Note that Equation~\ref{eq:bias} can be also expressed in terms of cross-correlations, which are commonly used to improve the signal-to-noise of the measurement. 

Figure~\ref{fig:spin_bias} shows the spin bias signal over 5 orders of magnitude in halo mass. Below a characteristic mass of $\log_{10} ({\rm M_{vir}}/h^{-1} {\rm M_{\odot}}) \simeq 11.5$, slower-rotating haloes 
are more tightly clustered than faster-rotating haloes of the same mass, an inversion produced by the presence of splashback haloes \citep{Tucci2020}. It is only above the aforementioned mass that the intrinsic spin bias signal, where higher-spin haloes have higher bias than lower-spin haloes, is observed. This signal is maximal for the most massive, largest haloes, reaching a factor $\sim$1.5 in the ratio of relative biases. As we will show in following sections, the SZ effects are expected to be stronger and easier to detect at the high-mas end.   

As shown in \cite{MonteroDorta2020B}, the spin bias signal measured from TNG300 differs slightly from the signal presented in Figure~\ref{fig:spin_bias} (and from other measurements using N-body numerical simulations). At the very low-mass end, the inversion produced by splashback haloes is not observed, whereas at the high-mass end, the signal vanishes for the most massive haloes (see Figure~\ref{fig:secondary} below). While the vanishing signal at the high-mass end is clearly caused by low-number statistics, the reasons for the lack of a low-mass inversion are still unclear and under investigation. These differences, however, do not affect the validity of our forecast, since the efficiency of the kSZ technique as a probe for spin bias is addressed {\it{relative}} to the amount of signal in TNG300. For discussion on the physical origins of the signal itself and its potential connection with the assembly bias trend, see \cite{SatoPolito2019, Johnson2019,Ramakrishnan2019,Mansfield2020,Tucci2020}.

\section{The observational probe}
\label{sec:kSZ}

\subsection{The tSZ and kSZ effects}
\label{sec:theory}

The SZ effect \citep{SZ1970,SZ1980A,SZ1980B} is the generation of temperature anisotropies in the observed CMB by the inverse Compton scattering of CMB photons as they propagate through galaxy clusters. In the {\it{thermal}} SZ effect,  the scattering of CMB photons is produced by the thermal energy of the electrons (random motion), and the magnitude of the signal is given by the dimensionless Compton parameter $y$:  

\begin{equation}
\label{eq:tSZ}
y(\vec{n}) \equiv  \frac{\sigma_T k_b}{m_e c^2} \int_{los} dl~T_e~n_e,  
\end{equation}

\noindent where $\vec{n}$ is the unit vector that defines the line of sight ($los$), $\sigma_T$ is the Thomson cross-section, $k_B$ is Boltzmann’s constant, $T_e$ is the electron temperature, $m_e$ is the electron rest mass, $c$ is the speed of light, and $n_e$ is the electron number density. The Compton parameter $y$ is therefore proportional to the electron pressure integrated along the $los$. This parameter is related to the relative change in temperature in the CMB, $\frac{\Delta T}{T_{\rm CMB}}$, in the following way:

\begin{equation}
\label{eq:conversion}
\left[\frac{\Delta T}{T_{\rm CMB}}\right]_{\rm tSZ} (\vec{n}) = g(x) y(\vec{n}) 
\end{equation}

\noindent where $g(x)=x{\text{coth}}(x/2)-4$ is the conversion factor at a given dimensionless frequency $x \equiv h\nu /(k_B T)$, and $T_{\rm CMB} = 2.725$ K \citep{Mather1999,Fixsen2009}. See, e.g., \cite{Rephaeli1995,Mroczkowski2019} for more details. 

The amplitude of the integrated tSZ effect is known to be a robust proxy for the total gas mass of the halo. The rotational properties of the gas, on the other hand, can be investigated using the second-order {\it{kinetic}} SZ effect. In the kSZ effect, the temperature fluctuations are caused by the Doppler shift due to bulk, coherent motion of the electrons. These fluctuations now depend on the integral along the $los$ of the density as well as the peculiar velocity of the electrons. Since the circumgalactic medium is optically thin to photons from the CMB, the relative change in temperature, $\frac{\Delta T}{T_{\rm CMB}}$, from free electrons in a halo can be expressed using the single-scattering limit, which also applies to the tSZ above. Namely:

\begin{equation}
\label{eq:kSZ}
\left[\frac{\Delta T}{T_{\rm CMB}}\right]_{\rm kSZ} (\vec{n}) = \frac{\sigma_T}{c} \int_{los} dl~n_e~\vec{v}~\cdot~\vec{n},  
\end{equation}

\noindent where $\vec{v}$ is the velocity of the electrons in the CMB rest frame. In contrast to the change in temperature due to the thermal effect, the kSZ is independent of frequency. Note, also, that the tSZ correlates with baryonic mass through $n_e$, whereas the kSZ is connected to both mass and rotational velocity, very much like angular momentum is.

A purely rotational motion of the intra-halo gas (i.e., as a rigid body) would produce temperature fluctuations in the form of a dipole on the plane of the detector (see, e.g. \citealt{Cooray2002, Chluba2002}), as long as the $los$ does not coincide with the rotation axis of the halo. In practice, the dynamics and geometry of the intra-halo gas are very complex and haloes are oriented in all possible directions. All these issues can debilitate the signal, but the power of the kSZ effect resides in the fact that it does not require an ideal orientation of the halo. The amplitude of the dipole depends on the integrated product of the $los$ component of the velocity $v_{los} = \vec{v} \cdot \vec{n}$ ~times the number density, so a certain amount of signal is, in most cases, produced (particularly for massive haloes). 

Finally, other SZ effects that are not addressed in this work are known to exist, including {\it{non-thermal}} SZ effects, {\it{relativistic corrections}} to the SZ effect, and {\it{polarised}} SZ effects, all arising from scattering of CMB photons with the free electrons residing in the potential wells of haloes. For a detailed description of the different SZ effects, their applications, and observability, see \cite{Mroczkowski2019}. For the theoretical derivation and physics behind the SZ effects we also refer the reader to, e.g., \cite{Rephaeli1995}.

\subsection{Measurement in IllustrisTNG}
\label{sec:method}

In order to measure the temperature anisotropies produced by the tSZ and the kSZ effects, the integrals in Equations~\ref{eq:tSZ} and~\ref{eq:kSZ} must be computed for the $\sim$50,000 haloes of TNG300. These computations are performed in a simplified but consistent manner in this work. First, all haloes are rotated so that the z-axis in the halo's reference frame corresponds to its rotation axis. Following \cite{Baldi2018}, the $los$ is oriented along the x-axis, i.e., always perpendicular to the rotation axis. This is of course unrealistic, but it allows us to measure the maximal ``intrinsic" signal. The effect of this choice is, in practice, small, so the main conclusions of this paper remain qualitatively the same when the rotation is not applied. Also for convenience, all haloes are assumed to be located at $z=0.05$. This distance is large enough that assuming that the $los$ is always parallel to the x-axis is a very good approximation.

The integration itself is performed dividing the space spanned by the halo in cubic integration cells that are significantly larger than the typical size of the gas cells, but small enough that this effective loss of resolution does not affect our results in any significant way. Within each of these cubic cells, the average values for the temperature (for tSZ) and the electron number density and $los$ velocity (kSZ) of the gas cells are taken. Throughout this work, we employ a grid of 100X100 cells on the plane of the detector (i.e., at $z=0$), independently of the size of the halo. This yields a varying resolution that increases for the smaller haloes. Although this choice is not common (or even possible) in observations, it provides us with a more precise measurement of the intrinsic SZ signals across the entire mass range. We have checked, nevertheless, that fixing the resolution to a small or to a very high value does not change our results qualitatively.

Each halo in the simulation is located at a particular position within the box and has a given group velocity associated to it. This peculiar velocity must be subtracted, since it would otherwise wash out the kSZ signal. We thus set the reference frame for each halo at its centre of mass (CM) and subtract the group position and velocity from the gas coordinates and velocities. 

The electron density of each gas cell within the halo, $n_e$, is obtained from the properties provided in the IllustrisTNG database. For non-star-forming gas cells, $n_e$ is simply computed as:
\begin{equation}\label{eq:ne}
    n_e = N_e \, X_H \, \frac{\rho_{\rm gas}}{m_{p}},
\end{equation}

\noindent where $N_e$ is the electron abundance, $\rho_{\rm gas}$ is the gas density of the cell, $X_H$ = 0.76 is the
hydrogen mass fraction, and  $m_{p}$ is the proton mass.

For star-forming cells, the sub-grid model adopted in IllustrisTNG assumes an unresolved multi-phase interstellar medium made of a mix of volume-filling warm gas and dense cold clumps, the latter containing most of the mass of the cell \citep{springel03}. Following the approach of \citet{Marinacci2017} and \citet{Stevens2019}, the ionised fraction of each gas cell is computed as the fraction of gas in the volume-filling warm component. In order to obtain this fraction, it is convenient to first determine the mass fraction of the cold gas phase, $f_{neutr}$, as \citep{springel03}:

\begin{equation}
    f_{neutr} = \frac{u_{\rm hot} - u}{u_{\rm hot} - u_{\rm cold}}\,,
\end{equation}

\noindent where $u$ is the internal energy per unit mass of the gas cells, with equation of state $u = P/[(\gamma-1) \rho]$, and $u_{\rm cold}$ and $u_{\rm hot}$ are the gas thermal energy of the cold and hot phases of each cell, respectively. The electron density of star-forming gas cells is simply computed from $f_{neutr}$ as $(1-f_{neutr})\,n_e$. We refer to \citet{Marinacci2017} and \citet{Stevens2019} for further details\footnote{In this work, we use the {\sc Dirty AstroPy} repository   (\url{https://github.com/arhstevens/Dirty-AstroPy}) from \citet{Stevens2019} to compute the ionised fraction of each gas cell.}.  

Finally, the gas temperature, which is necessary to measure the tSZ effect through Equation~\ref{eq:tSZ}, is computed using the mass, internal energy, and electron abundance of each individual cell.

The main results of this paper are obtained with the TNG300 box, which is large enough to ensure a good clustering measurement. In the Appendix, we provide a comparison with the smaller but higher-resolution TNG100 box.

\begin{figure*}
\includegraphics[width=2\columnwidth]{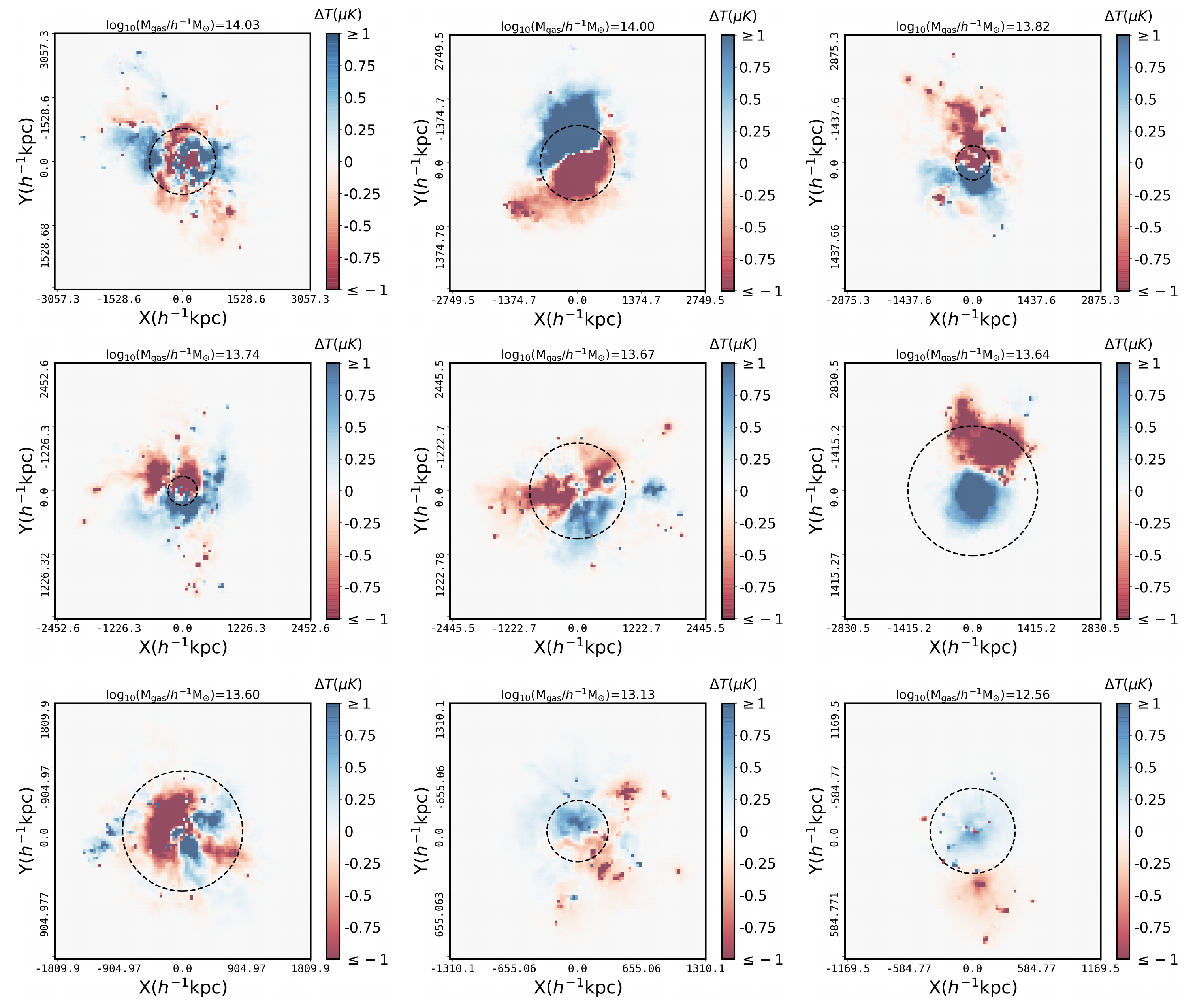}
\caption{kSZ maps showing $[\Delta T]_{\rm kSZ}$ for 9 randomly-selected haloes in TNG300, with gas masses in the range $12.5 \lesssim \log_{10} ({\rm M_{gas}}/h^{-1} {\rm M_{\odot}}) \lesssim 14$. In each panel, the kSZ temperature fluctuations in units of $\mu$K, computed from Equation~\ref{eq:kSZ} assuming $T_{\rm CMB} = 2.725$ $K$, are shown for each pixel on the $x'-y'$ plane. Note that, in order to emphasise the effect, haloes have been rotated so that the $los$ is placed in the direction perpendicular to the rotation axis of each halo. The {\it{field of view}} for each halo is determined here by the maximum separation of the gas particles with respect to the halo centre on the $x'-y'$ plane (L$_{\rm max}$), with resolution equal to (2$\times$L$_{\rm max})/100$. The halo-centric circle indicates the region where 2/3 of gas particles reside. Finally, in order to show the kSZ substructure of the haloes, saturation limits at $[\Delta T]_{\rm kSZ} = \pm 1$~$\mu K$ are set in the colour bar. This ensures that the maps are not dominated by a few pixels. Note that the fluctuations can easily reach a few dozen $\mu K$, but this peak signal is only observed in a reduced number of pixels.}
\label{fig:maps}
\end{figure*}

\begin{figure*}
\includegraphics[width=2\columnwidth]{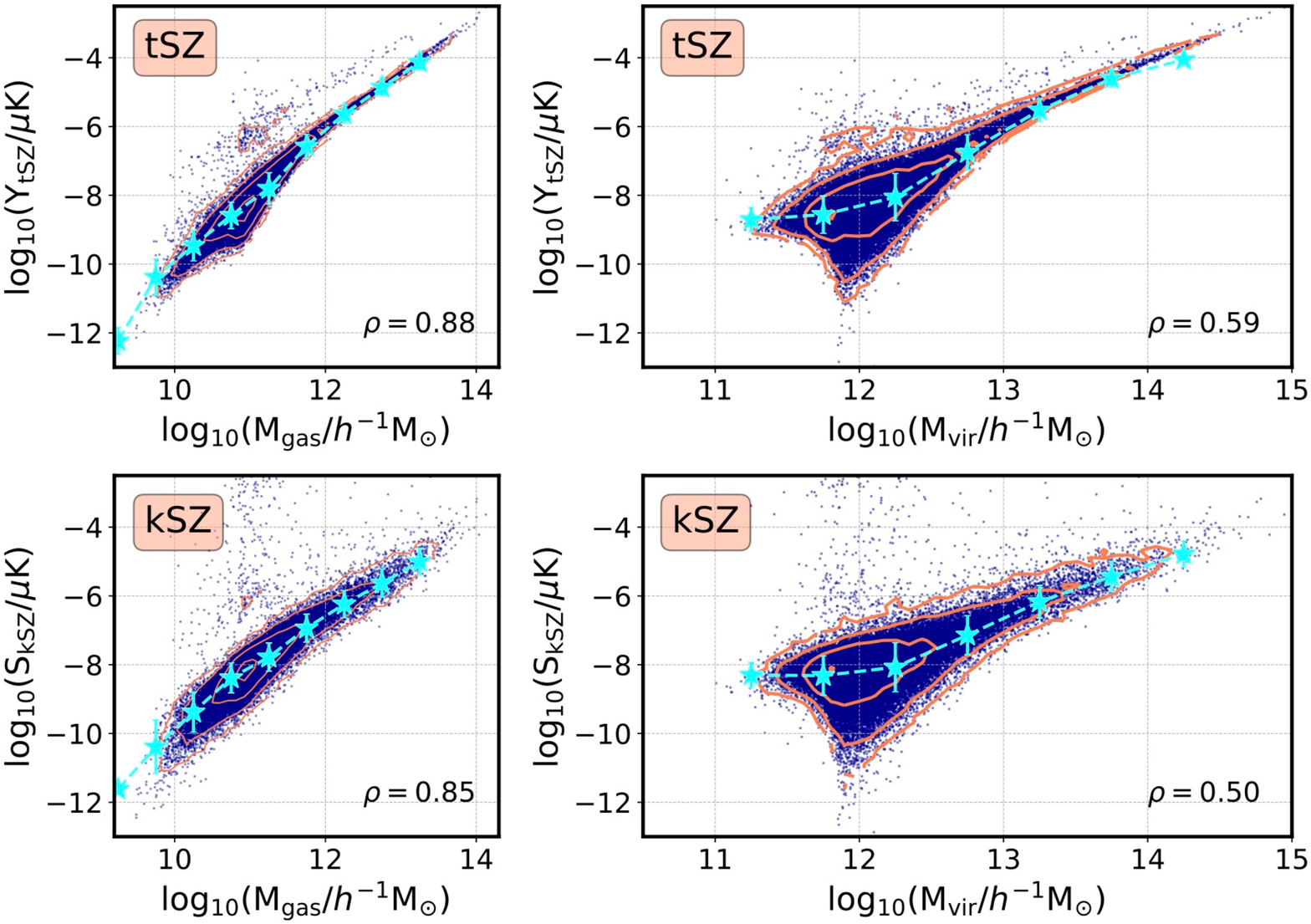}
\caption{{\it{Upper row:}} The correlation between the integrated tSZ signal, Y$_{\rm tSZ}$  (expressed in logarithmic units, see Equation~\ref{eq:tSZ_signal}), and the logarithm of the total gas mass (left) and the virial mass of the halo (right), respectively. Mean values and $\pm$1-$\sigma$ uncertainties are shown on top of the data points. {\it{Lower row:}} The same for the integrated kSZ signal, S$_{\rm kSZ}$ (see Equation~\ref{eq:kSZ_signal}). The Spearman rank-order correlation coefficient, $\rho$, is provided in the bottom-right corner of each panel.}
\label{fig:correlations_mass}
\end{figure*}

\subsection{The tSZ and kSZ statistics}
\label{sec:signal}

By computing Equations~$\ref{eq:tSZ}$ and $\ref{eq:kSZ}$ in pixels on the plane perpendicular to the $los$ at $z=0$ (i.e, the detector's plane, $x'-y'$), tSZ and kSZ 2-D maps can be obtained. The tSZ, in particular, produces a single-peak distribution centred at the halo's CM. From the Compton $y$-maps, a simple statistic, the integrated tSZ signal, Y$_{\rm tSZ}$, is defined for each halo as the sum of all pixels. Namely: 

\begin{equation}
   {\rm Y_{tSZ}} = D ~{\rm T_{CMB}} \sum_{\rm map}~y(x',y'),
   \label{eq:tSZ_signal}
\end{equation}

\noindent where the sum extends to the entire map and $D$ is the ratio of the area of the pixel in physical units and the square of the angular distance to $z=0.05$, i.e., $D = s^2/d_A^2$. The area $s^2$ depends on the particular halo, since the resolution varies. The angular distance $d_A$, however, is assumed to be the same for all haloes. The CMB temperature is introduced for consistency with the kSZ statistic below, and $T_{\rm CMB} = 2.725$ K is assumed \citep{Mather1999,Fixsen2009}. Note that, for simplicity, we are implicitly disregarding the conversion factor in Equation~\ref{eq:conversion}.

The definition of an integrated signal is a bit trickier for the kSZ effect, which produces a dipole pattern on the plane of the detector. In order to compress the kSZ information on the kSZ maps into a single statistic, the integrated kSZ dipole signal, S$_{\rm kSZ}$, is defined in terms of a simple sum of pixel values above and below a chosen demarcation. This demarcation, which separates the dipole components, is defined by the angle, $\theta$, with respect to the $x'$-axis on the plane of the detector. The signal S$_{\rm kSZ}^{\theta}$, for a given $\theta$, can therefore be expressed as:

\begin{equation}
   {\rm S_{kSZ}^{\theta}} = D \mathlarger{\mathlarger{\sum_{+}}}~\left[\Delta T \right]_{\rm kSZ}  - D \mathlarger{\mathlarger{\sum_{-}}}~\left[\Delta T \right]_{\rm kSZ},
   \label{eq:kSZ_signal}
\end{equation}

\noindent where the suffixes ``+" and ``-" in the summation terms refer to the two halves, above and below the demarcation. Note that this choice of statistic is completely arbitrary, and, in fact, the statistic could be defined differently in a real observation. As shown in following sections, we have simply chosen the one that provides a tighter correlation with $\lambda_{halo}$ (and with the mass and angular momentum of the gas), while being easy enough to compute for a large sample of haloes. 

\begin{figure*}
\includegraphics[width=2\columnwidth]{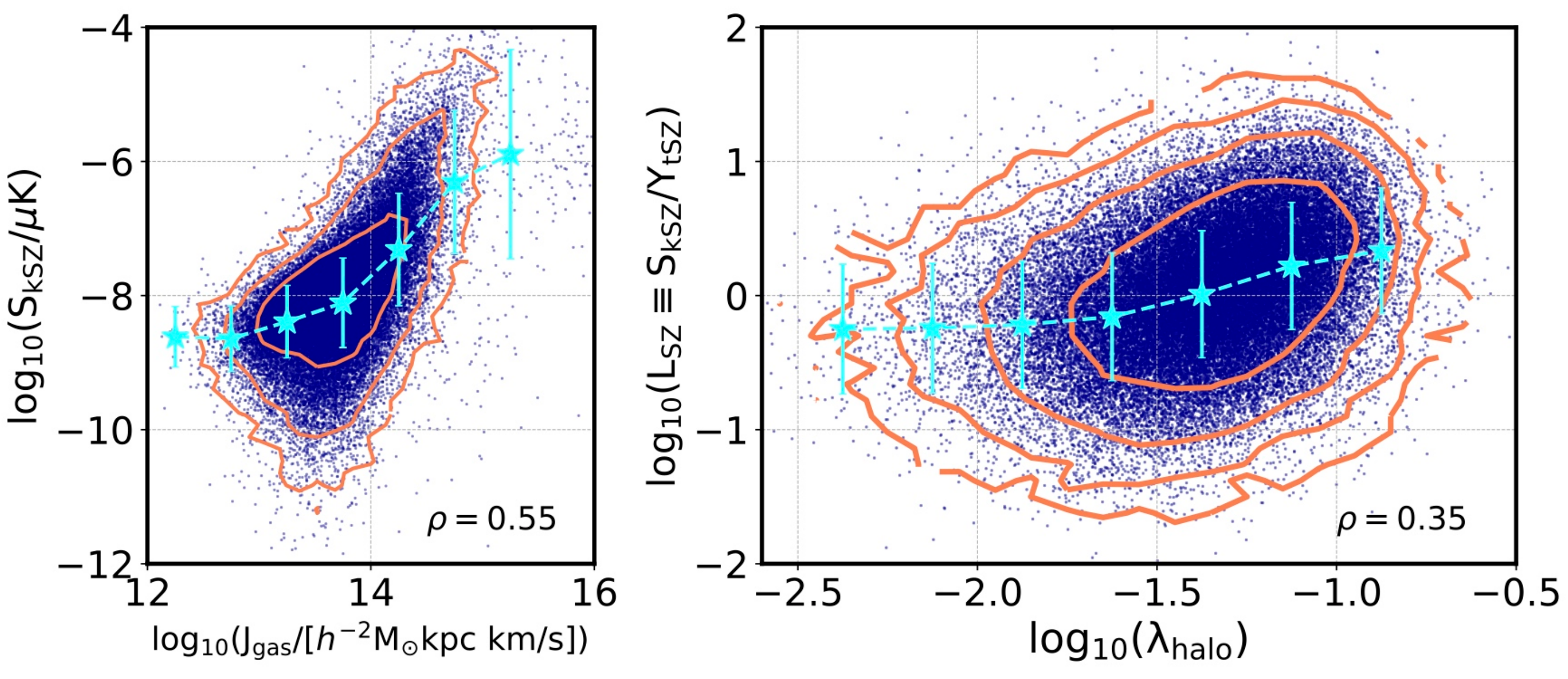}
\caption{{\it{Left:}} The correlation between the integrated kSZ signal, S$_{\rm kSZ}$ (expressed in logarithmic units, see Equation~\ref{eq:kSZ_signal}), and the logarithm of the total angular momentum of the gas. Mean values and $\pm$1-$\sigma$ uncertainties are shown on top of the data points. {\it{Right:}} In the same format, the ratio of the integrated kSZ and tSZ signals, L$_{\rm SZ}$ (see Equation~\ref{eq:SZ_signal}), as a function of halo spin, in logarithmic units. The Spearman rank-order correlation coefficient, $\rho$, is provided in the bottom-right corner of each panel.}
\label{fig:correlations_spin}
\end{figure*} 
 
The ideal configuration for the kSZ depends on the particular orientation of the halo on the plane of the detector. Since we analyse a large sample in this work (comprising more than 50,000 haloes), we must automatise the placement of the dividing demarcation, which should follow the line separating the dipole components. In order to do so, we define 50 different orientations for the demarcation, $\theta_i(deg)$ $\in$ [-90,90], which cover the entire range of choices. The final statistic that we use throughout this work, which we simple denote by S$_{\rm kSZ}$, corresponds to the configuration that maximises the signal.

The practical implementation of the above method faces several problems. First, the maps contain ourliers, i.e., pixels for which the signal is unrealistically large. These outliers are produced by small-scale substructures that occasionally have very large projected velocities (see, e.g., \citealt{Baldi2018}). Since they are mostly uncorrelated with the rotational properties of the haloes, they need to to be masked out. We alleviate this problem by removing all pixels with kSZ signals above a 5-$\sigma$ level from the mean (for each half of the detector). Second, for a small fraction of haloes, the dipole is not centred on the projection of the halo's CM on the detector's plane. In order to find the centre, we simply compute the module of the gradient of the signal on the plane, and smooth it with a Gaussian kernel of a few pixels width. By doing this, we can identify the region on the map where the change in the temperature fluctuations is larger, which corresponds to the line that separates the dipole components. The Gaussian smoothing ensures that this is not dominated by any outliers.

Finally, it is convenient to define the ratio between the kSZ and the tSZ signals, which we call L$_{\rm SZ}$:

\begin{equation}
   {\rm L_{\rm SZ}} = {\rm \frac{S_{kSZ}}{Y_{tSZ}}}.
   \label{eq:SZ_signal}
\end{equation}

Note that using the L$_{\rm SZ}$ dimensionless statistic in the context of this work is motivated by the Bullock et al. definition of halo spin, which can be expressed in terms of the {\it{specific}} angular momentum of the halo, i.e., the angular momentum per unit virial mass (Equation~\ref{eq:spin}).  

\section{Results}
\label{sec:results}

In this section, we provide examples of kSZ maps, discuss the scaling relations with respect to several halo properties of interest, and analyse the clustering of haloes as a function of the S$_{\rm kSZ}$ and L$_{\rm SZ}$ statistics. 

\subsection{The kSZ maps}
\label{sec:maps}

Figure~\ref{fig:maps} displays some examples of 2D kSZ maps, $[\Delta T]_{\rm kSZ} (x',y')$, measured on the plane of the detector, $x'-y'$. Nine randomly-selected haloes with gas contents expanding the range $12.5 \lesssim \log_{10} ({\rm M_{gas}}/h^{-1} {\rm M_{\odot}}) \lesssim 14$ are shown. These haloes have been rotated so that the $los$ lies perpendicular to their rotation axis, a choice that, by definition, should maximise the amplitude of the dipole pattern\footnote{In practice, due to the turbulent motion of gas inside haloes, the effect is relatively small.} (which is proportional to the integrated product $\vec{v}~\cdot~\vec{n}$).

In each panel of Figure~\ref{fig:maps}, the {\it{field of view}} expands twice the projected maximum separation of a gas particle with respect to the projection of the halo's CM on the $x'-y'$ plane (2*L$_{\rm max}$). For the representative sample displayed in Figure~\ref{fig:maps}, L$_{\rm max}\sim$1-3 $h^{-1}$Mpc. We employ 100X100 pixels in our calculation, a configuration that yields a typical resolution between 20X20 and 60X60 $h^{-2}$kpc$^{2}$, approximately. A scale of 20 kpc correspond to $\sim$ 20 arcsec at $z=0.05$. As a reference, the recently claimed detection of the kSZ effect in the cluster system MACS J0717.5+3745 is based on a measurement that reaches an effective angular resolution of 22 arcsec \citep{Adam2017}. Note that these sub-arcmin resolutions are currently very challenging to achieve, but they will be attainable with future instrumentation for large samples of clusters (see Section~\ref{sec:discussion} below). Importantly, we have checked that degrading the resolution in our computation does not have a qualitative impact on our results.   

Figure~\ref{fig:maps} illustrates some of the typical cases that we encounter in our sample, with dipoles being more evident for some haloes than others. In order to provide a better visualisation of the halo structure, a ``saturation limit" in the colour bar of our maps is imposed (i.e., pixels with absolute values above these limits are painted in the same colour). This is necessary, since the signal is strongly dominated by the inner regions of the haloes, where the majority of the particles reside (the circles represent the region around the CM containing 2/3 of the particles). In these regions, the peak signal can reach dozens of $\mu K$ in $[\Delta T]_{\rm kSZ}$ for the largest haloes. 

In most cases, the demarcation separating the two components of the dipole is fairly obvious. However, as expected, we also find cases where the turbulent motion of gas inside the halo prevents the formation of a clear dipole (top left panel of Figure~\ref{fig:maps}). The halo at the bottom-right corner of the figure is also problematic, since the dipole is not centred at the CM of the halo. As described in the previous section, we have developed a simple and efficient way to correct this issue.

\subsection{The scaling relations}
\label{sec:scaling}

Although, for simplicity, we opt to show only kSZ maps, tSZ $y(x',y')$-maps are also computed for all haloes in TNG300. From these two sets of maps, the three statistics defined in Section~\ref{sec:signal}, i.e., S$_{\rm kSZ}$, Y$_{\rm tSZ}$, and L$_{\rm SZ}$, are measured for each halo. In this section, we evaluate the performance of these quantities as tracers of several gas and virial halo properties. 

We begin by analysing the mass components in  Figure~\ref{fig:correlations_mass}. This figure demonstrates that, 1) both the tSZ and the kSZ statistics can be considered good tracers of the total gas mass and virial mass of the halo (which is dominated by the DM component), and, 2) the integrated tSZ signal provides significantly better results at the high-mass end. In the upper row, the logarithm of Y$_{\rm tSZ}$ is shown as a function of gas mass (left) and virial mass (right); the same relations are displayed for S$_{\rm kSZ}$ in the lower row. Although the two SZ statistics naturally differ in amplitude (this is irrelevant for our analysis), both exhibit similarly-shaped correlations with the two mass quantities under analysis, with a scatter that increases towards the low-mass end. The scatter in the Y$_{\rm tSZ}$--mass relations is, however, smaller, becoming almost imperceptible for the most massive haloes, where the performance of the tSZ proxy is outstanding. As we explain below, the larger scatter in the S$_{\rm kSZ}$--mass relation, particularly at the high-mass end, is a consequence of the kSZ tracing the rotational properties of haloes at fixed halo mass. The Spearman rank-order correlation coefficient, $\rho$, shown in the bottom-right corner of the plots, is, as expected, higher for the thermal effect. 

The potential of the kSZ technique as an observational probe for spin bias depends mostly on how well the integrated kSZ signal, S$_{\rm kSZ}$, traces the rotation of the intra-halo gas and how this is related to the total spin of the halo. The left-hand panel of Figure~\ref{fig:correlations_spin} displays the logarithm of S$_{\rm kSZ}$ as a function of the logarithm of the total angular momentum of the intra-halo gas, J$_{\rm gas}$. The correlation is again remarkable, although the scatter becomes large at the low-mass end, due to the turbulent motion of gas inside the haloes (see, e.g., \citealt{Battaglia2012, Yu2015}). The correlation coefficient for the entire mass range is still quite good, $\rho = 0.55$. 

Halo spin is proportional to the {\it{specific}} angular momentum, $j$, the angular momentum per unit mass (Equation~\ref{eq:spin}). As shown above, the kSZ signal is proportional to the total angular momentum and, therefore, also to the mass of the intra-halo gas. For this reason, halo spin should be more directly connected to the ratio of the integrated kSZ ($\propto$ J, M) and tSZ signals ($\propto$ M), which are computed independently. The right-hand panel of Figure~\ref{fig:correlations_spin} displays the relation between L$_{\rm SZ}$ and halo spin in TNG300. As expected, the scatter increases with respect to the left-hand panel of the same figure, since halo spin also includes the DM component of the halo. A correlation is nevertheless evident, with a coefficient of $\rho = 0.35$. Note that here, the entire range of halo masses is considered,  decreases the correlation with $\lambda_{\rm halo}$ (as opposed to restricting the analysis to the high-mass end). 

Although, theoretically, L$_{\rm SZ}$ should be more directly linked to $\lambda_{\rm halo}$, we have checked that S$_{\rm kSZ}$ displays just a marginally worse correlation with $\lambda_{\rm halo}$, i.e., $\rho = 0.33$. In practice, by dividing by Y$_{\rm tSZ}$, we might introduce an extra amount of scatter (particularly at the low-mass end) that could eliminate the potential advantages of using L$_{\rm SZ}$.   

The scatter found between L$_{\rm SZ}$ (or S$_{\rm kSZ}$) and $\lambda_{\rm halo}$ is, in any case, not surprising, given the complex dynamics and composition of the intra-halo gas. As we show in the next section, this does not imply that the spin bias signal cannot be recovered, at least partially, using L$_{\rm SZ}$ or S$_{\rm kSZ}$ as a halo discriminator. The clustering measurements are statistical in nature and often a strong correlation is not required for the signal to be ``transmitted" (see discussion on halo/galaxy spin bias in \citealt{MonteroDorta2020B}).

\begin{figure}
\includegraphics[width=0.98\columnwidth]{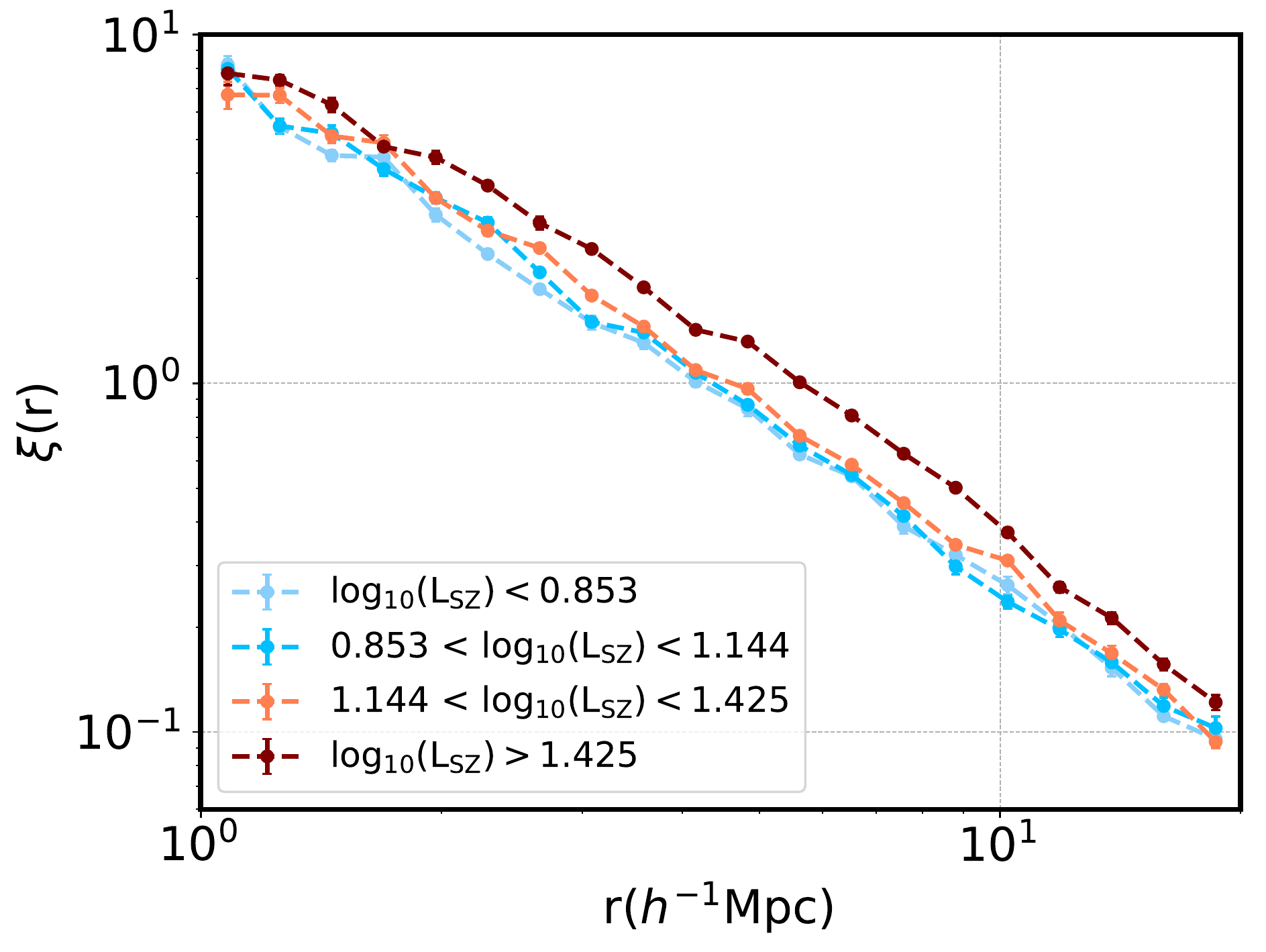}
\caption{The two-point correlation function, $\xi$, expressed as a function of 3D separation $r$, for the four quartiles defined in terms of the logarithm of the ratio of the integrated kSZ and the tSZ signals, L$_{\rm SZ}$, measured from TNG300. A log-spaced binning is assumed. The uncertainties, which are almost indistinguishable in this range of scales, are obtained from a set of jackknife subsets.}
\label{fig:CF}
\end{figure}

It is worthwhile to stress once more the differences between angular momentum and spin. The angular momentum depends on mass and, therefore, it indirectly correlates with the  integrated tSZ signal Y$_{\rm tSZ}$ ($\rho = 0.45$). However, the Bullock et al. definition of spin eliminates the mass dependence, thus isolating the rotational effect. We have checked that Y$_{\rm tSZ}$ displays almost null correlation with $\lambda_{\rm halo}$ (a marginal $\rho = 0.08$), which in fact demonstrates that the correlation with angular momentum comes from the mass itself. By comparing Figures~\ref{fig:correlations_mass} and~\ref{fig:correlations_spin}, one can realise that the larger scatter observed between S$_{\rm kSZ}$ and M$_{\rm vir}$ (as compared to Y$_{\rm tSZ}$--M$_{\rm vir}$) is reflecting the secondary dependence on the rotational properties of haloes, which manifests itself in the correlation with spin.

We have checked that other possible definitions of S$_{\rm kSZ}$, Y$_{\rm tSZ}$, and L$_{\rm SZ}$, including those based on the mean/median instead of the sum, or on restricting the calculation to the inner regions of the haloes, produce also correlations with the quantities considered. Our choice yields the best results in the context of this analysis with the IllustrisTNG simulation, but the situation could be different in other contexts. Finally, in the Appendix, we provide a comparison between the signals and scaling relations found for the TNG300 and TNG100 boxes. As explained in Section~\ref{sec:sims},  TNG100 is significantly smaller than TNG300, which makes it unsuitable for clustering studies. It provides, however, much higher resolution. In the Appendix, we demonstrate that our results are not affected by the lower resolution of TNG300. 

\subsection{Clustering measurements}
\label{sec:clustering}

\begin{figure*}
\includegraphics[width=1.7\columnwidth]{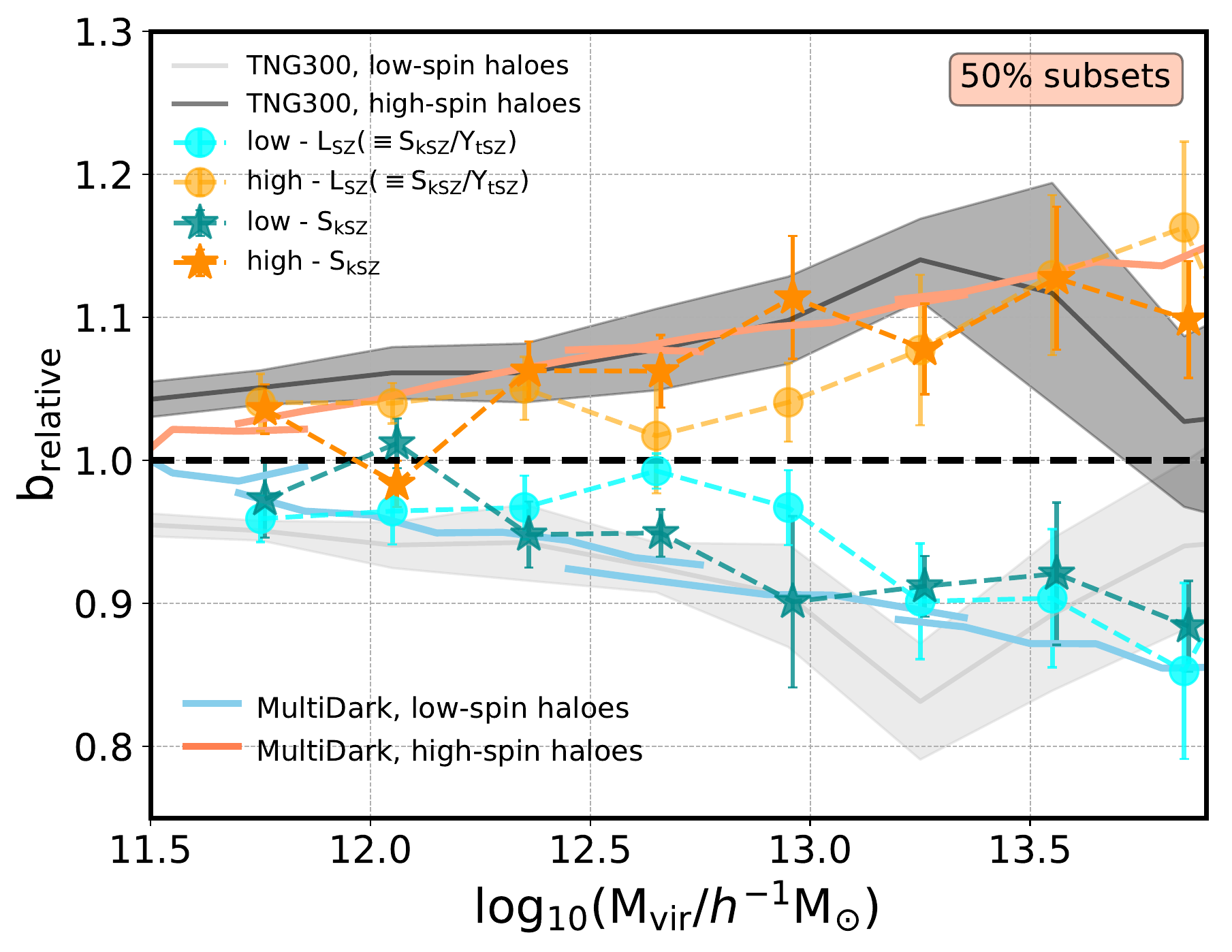}
\caption{The secondary dependence of halo bias on the integrated kSZ signal and the ratio of the integrated kSZ and tSZ signals, at fixed halo mass. The dark cyan and dark orange stars show, respectively, the relative bias for the 50$\%$ halo subpopulations with lower and higher S$_{\rm kSZ}$ (the statistic based on the kSZ effect alone). The light cyan and light orange circles represent the same type of subsets selected in terms of L$_{\rm SZ}$ (the ratio of statistics). Error bars on these measurements correspond to the jackknife uncertainties explained in the text. The SZ clustering measurements are compared with two halo spin bias measurements: the underlying TNG300 singal (grey lines) and the MultiDark (N-body simulation) results (coral/blue lines). Note that the S$_{\rm kSZ}$ data points have been shifted slightly along the x-axis in order to make the error bars distinguishable.}
\label{fig:secondary}
\end{figure*}

TNG300, with 205 $h^{-1}$Mpc of side length, is one of the largest hydrodynamical simulation boxes available to the community and one of the first that actually allows for some statistically-significant measurements of clustering (for certain scales and mass ranges, see, e.g., \citealt{MonteroDorta2020B}). In this section, we use the spatial distribution of haloes in TNG300 to test whether the halo spin bias signal can be recovered from measurements of the SZ effects (the S$_{\rm kSZ}$, Y$_{\rm tSZ}$, and L$_{\rm SZ}$ statistics). 

Following the standard procedure in simulations (see Section~\ref{sec:spin_bias}), we use the 3D two-point correlation function, $\xi(r)$, to measure the relative clustering between different subsets of SZ-selected haloes. In Figure~\ref{fig:CF},  $\xi(r)$ is plotted on scales $1 \leq r[h^{-1}\rm{Mpc}] \leq 20$ as a function of the logarithm of L$_{\rm SZ}$. The full sample of TNG300 haloes is here divided in L$_{\rm SZ}$ - quartiles, i.e., subsets of haloes defined using L$_{\rm SZ}$ and containing 25$\%$ of the entire population. The computation of errors is based on a standard jackknife technique, where a portion of 1/8 the total volume of the box is removed each time (see below). Figure~\ref{fig:CF} displays a quite stable $\xi(r)$ measurement within the range of scales considered, along with an interesting dependence on L$_{\rm SZ}$. The amplitude of clustering tends to increase with L$_{\rm SZ}$, but only the highest quartile separates from the rest significantly. A similar behaviour is found for S$_{\rm kSZ}$.

Importantly, since the primary determinant of halo clustering is halo mass, the implicit distribution of halo masses in each subset shown in Figure~\ref{fig:CF} has a strong effect on the trend. In order to better isolate the dependence on the S$_{\rm kSZ}$ and L$_{\rm SZ}$ statistics, we measure the {\it{relative bias at fixed halo mass}}, following the procedure developed in \cite{MonteroDorta2020B} for TNG300. The key aspect of this computation is the use of the cross-correlation between quartiles and the entire halo sample, which maximises the signal-to-noise of the measurement (this is important given the small size of TNG300). For a given halo-mass bin M$_{i}$ and subset $\mathcal{S}$ (defined in terms of S$_{\rm kSZ}$ or L$_{\rm SZ}$), the relative bias can be measured as:

\begin{equation}
   {\rm b}_{\rm relative}({\rm r},\mathcal{S}|{\rm M_i}) = \frac{\xi_{[\mathcal{S},all]}({\rm r})}{\xi_{[{\rm M_{i}},all]}({\rm r})},
   \label{eq:bias2}
\end{equation}

\noindent where $\xi_{[S,all]}$ is the cross-correlation between all objects in the subset and all objects in the sample, and $\xi_{[{\rm M_i},all]}$ is the cross-correlation between all objects in the halo-mass bin and the entire sample as well. The estimation of the relative bias is performed by averaging over the range of scales 5 $ < r[h^{-1}{\rm{Mpc}}] <$ 12 (here, the upper limit is lowered slightly, in order to avoid problems related to the small size of the box, see \citealt{MonteroDorta2020B}). Note that the relative bias can be computed either from the auto-correlations (Equation~\ref{eq:bias}) or the cross-correlations (Equation~\ref{eq:bias2}). The computation of errors is based on a standard jackknife technique, where the TNG300 box is divided in 8 sub-boxes ($ L_{sub-box} = L_{box}/2 = 102.5$ $h^{-1}$Mpc). The relative bias is measured in 8 different configurations of equal volume, obtained by subtracting one sub-box at a time. The errors on the measured relative bias correspond to the standard deviation of all individual configurations. 

In Figure~\ref{fig:secondary}, the secondary bias, b$_{\rm relative}$, is displayed for the two 50$\%$ L$_{\rm SZ}$-subsets and the two 50$\%$ S$_{\rm kSZ}$-subsets (i.e., taking for both statistics half the population above and below the median). These measurements are compared with the TNG300 and MultiDark halo spin bias signals in the halo mass range $11.5 < \log_{10} ({\rm M_{vir}}/h^{-1} {\rm M_{\odot}}) < 13.7$. In order to make measurements fully comparable, the MultiDark relative-bias results are here obtained using 50$\%$ subsets as well, instead of the standard 25$\%$ quartiles that only the MultiDark statistics permit. The first important result of Figure~\ref{fig:secondary} is the excellent agreement between the spin bias found in MultiDark and TNG300 (except for the very-high-mass end, where the TNG300 signal vanishes due to the lack of statistics). This is very reassuring in terms of the robustness of secondary-bias measurements in TNG300. 

Figure~\ref{fig:secondary} also shows a systematic difference in relative bias between subsets defined both in terms of S$_{\rm kSZ}$ and L$_{\rm SZ}$, which traces the spin bias trend. Interestingly,
despite the fact that L$_{\rm SZ}$ provides a slightly better  correlation with $\lambda_{\rm halo}$, it is S$_{\rm kSZ}$ that more closely follows the underlying $\lambda_{\rm halo}$ trend (note that the L$_{\rm SZ}$ data points are, nevertheless, statistically compatible). It is of course at the high-mass end that we find the largest separations in b$_{\rm relative}$ for both statistics. The spin bias signal is small at the low-mass end, but it reaches around 15$\%$ at $\log_{10} ({\rm M_{\rm vir}}/h^{-1} {\rm M_{\odot}}) = 13$ and $\sim$25$\%$ at $\log_{10} ({\rm M_{\rm vir}}/h^{-1} {\rm M_{\odot}}) \simeq 14$. Figure~\ref{fig:secondary} shows that the performance of the two SZ-based statistics is very good {\it{relative to the underlying signal}}. Importantly, by dividing by the integrated tSZ signal in L$_{\rm SZ}$, we are mitigating the potential effect of halo mass in Figure~\ref{fig:secondary}.

In order to provide more insight on the correlation between the SZ-based statistics and $\lambda_{\rm halo}$, we have measured the overlap between subsets selected using these quantities, following \cite{MonteroDorta2020B}. The {\it{average subset overlap}} is defined as the average percentage of objects that pertain to the high/low $\lambda_{\rm halo}$-subset that are also part of the high/low L$_{\rm SZ}$- (or S$_{\rm kSZ}$-) subset\footnote{Note that this is basically a simpler, more intuitive alternative to the Spearman rank-order correlation coefficient, $\rho$.} (the overlap is measured between corresponding subsets and the average between the two resulting values is taken). Note that a value around 50$\%$ would indicate that the SZ-based selection is not biased towards any particular type of haloes, which would imply little correlation. This is clearly not the case, as our analysis yields an overlap of around 60-70$\%$ across the halo mass range considered for both statistics.  

\begin{figure}
\includegraphics[width=1\columnwidth]{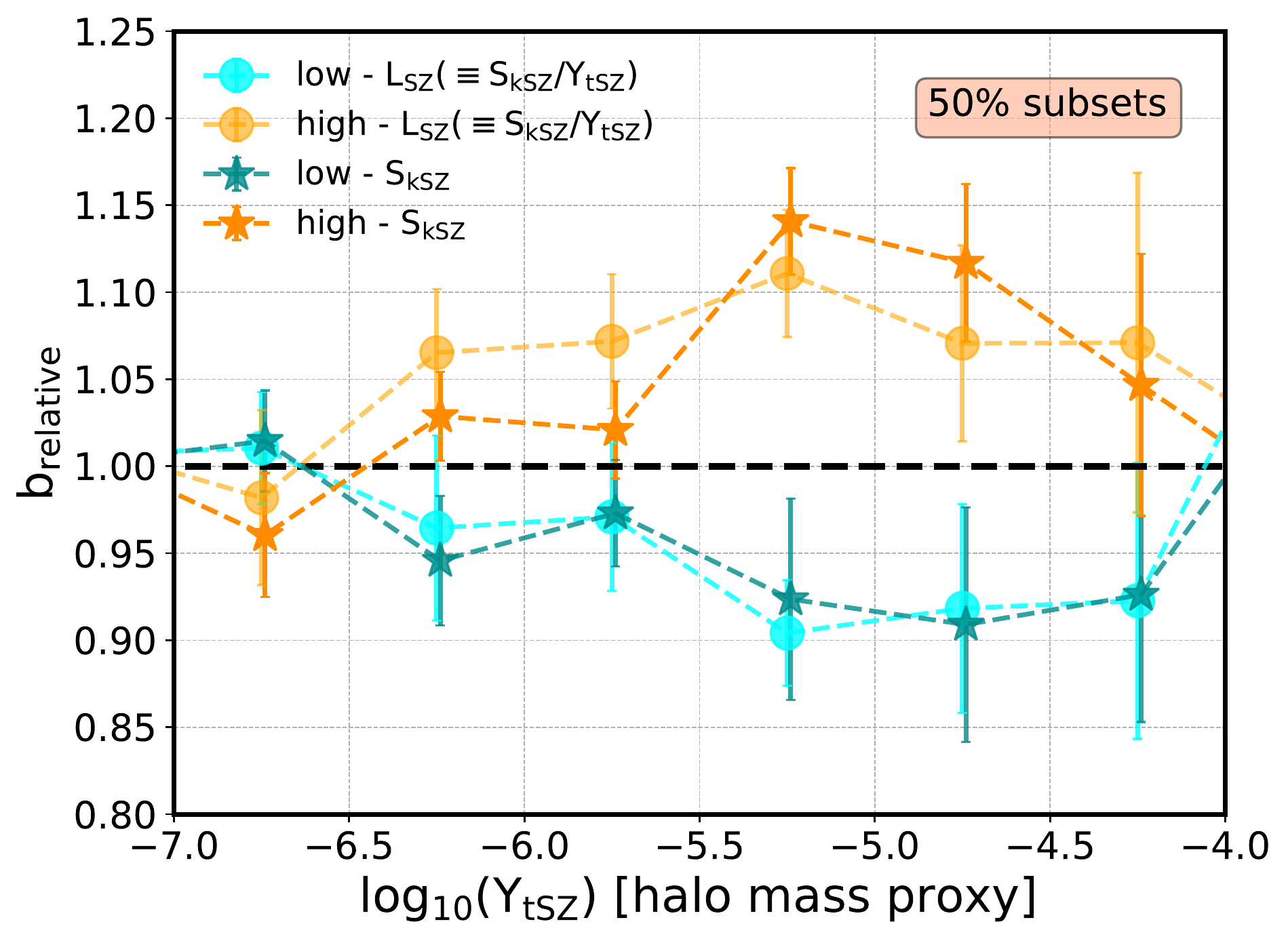}
\caption{The secondary dependence of halo bias on both the integrated kSZ signal, S$_{\rm kSZ}$, and the ratio of the integrated kSZ and tSZ signals, L$_{\rm SZ}$, in bins of the tSZ statistic, Y$_{\rm tSZ}$. Both S$_{\rm kSZ}$ and L$_{\rm SZ}$ are proxies for halo spin, whereas Y$_{\rm tSZ}$ is a proxy for halo (virial) mass. The dark cyan and dark orange stars show the relative bias for low- and high-S$_{\rm kSZ}$ haloes, whereas the light cyan and light orange circles represent the same type of subsets selected in terms of L$_{\rm SZ}$ (50\% subsets). Uncertainties on these measurements correspond to the jackknife errors described in the text. Note that the S$_{\rm kSZ}$ data points have been shifted slightly along the x-axis in order to make the error bars distinguishable.}
\label{fig:secondary_tSZ}
\end{figure}

We now go a step further and test the inclusion of Y$_{\rm tSZ}$ as a proxy for halo mass in Figure~\ref{fig:secondary} (x-axis). Note that the uncertainties in the determination of halo masses are a major obstacle for probing secondary bias, since the effects must be measured at {\it{fixed halo mass}}. This, given the magnitude of the secondary bias signals, implies a precision in the mass measurement of at least $\sim0.2$ dex \citep{SatoPolito2019}. In Figure~\ref{fig:secondary_tSZ}, we show the relative bias between haloes selected by L$_{\rm SZ}$ and S$_{\rm kSZ}$ (proxies for halo spin) as a function of Y$_{\rm tSZ}$ (proxy for halo mass). We restrict this analysis to haloes with tSZ signals Y$_{\rm tSZ}>-7$, which typically correspond to halo masses $\log_{10} ({\rm M_{vir}}/h^{-1} {\rm M_{\odot}}) > 12.5$. Below this threshold, the underlying spin bias signal is too small and the scatter in the Y$_{\rm tSZ}$--M$_{vir}$ relation becomes large. Figure~\ref{fig:secondary_tSZ} illustrates the potential of the tSZ effect as a halo mass proxy in the context of secondary bias, and spin bias in particular. As expected from the small scatter in the Y$_{\rm tSZ}$--M$_{\rm vir}$ relation at the high-mass end, spin bias continues manifesting itself when Y$_{\rm tSZ}$ is used to bin the data. The separation in relative bias is statistically significant, even though, again, the error bars are dominated by the small cosmological volume of TNG300. 

\section{A note on the current state of SZ observations}
\label{sec:discussion}

Our analysis shows the potential of the kSZ effect, in combination with the tSZ effect, as a future observational probe for spin bias. This idea adds to previously discussed cosmological applications, such as the mapping of the large-scale motions of baryons in the Universe, constraining the growth of structure or testing isotropy/homogeneity in the Universe (e.g., \citealt{HM2006,Bhattacharya2007,Bhattacharya2008,Zhang2011,Hand2012, Eva2015, Alonso2016}). It is important to bear in mind, however, that measuring the kSZ signal for a large number of clusters is not possible with current instrumentation, unless a statistical (``stacking") approach is employed \citep{Hand2012,Li_kSZ2018}.

A more realistic assessment of the detectability of the effect would require simulating the observational and instrumental uncertainties that currently beset individual measurements. In this context, one of the main problems for measuring the resolved kSZ effect on an object-by-object basis is that the signal follows the same spectral dependence as the CMB. It can only be identified by measuring a dipole on the plane of the detector, which requires both high sensitivity and high angular resolution. In addition to this technical challenge, the kSZ signal is intrinsically small (approximately an order of magnitude weaker than the tSZ effect) and thus hard to disentangle from a variety of other astrophysical signals, including dust and radio emission (see discussion in, e.g., \citealt{Sayers2019,Mroczkowski2019}). Simulating the {\it{observed}} ~kSZ signal for a given experiment would require mimicking its cluster selection pipeline and accounting for the CMB sky (with associated experimental noise) in order to include all possible systematics.  

Despite the aforementioned challenges, significant progress has been made in the last years, which lead to tentative detections of the kSZ signal in a handful of clusters (see, e.g., \citealt{Mroczkowski2012,Sayers2013,Schaan2016,Adam2017,Sayers2019}). These measurements have proven effective for determining the peculiar velocities of the clusters with respect to the CMB. 

The landscape is significantly more clear form tSZ observations. Since the first single-object detections by \cite{Birkinshaw1984}, the number of tSZ clusters available has increased steadily. In recent years, the largest samples have been compiled by survey instruments operating at millimetre wavelengths, including the Atacama Cosmology Telescope (ACT, \citealt{Hasselfield2013}), the South Pole Telescope (SPT, \citealt{Bleem2015}), and the Planck satellite \citep{Planck_kSZ2016}. With spatial resolutions that range from 1.5 arcmin to 10 arcmin (at the wavelengths relevant for the detection of the SZ effect), they have delivered samples of 91, 747 and 1963 tSZ clusters, respectively. In addition, the Planck Collaboration has recently reported the first all-sky Compton-y map \citep{Planck2013}. These data sets have served as the basis for a number of cosmological parameter determination analyses (see recent examples in \citealt{Hurier2017,Bolliet2018, Bocquet2019, Bolliet2020}). 

It is projected that upcoming ground-based instrumentation reaches resolutions in the  sub-arcmin range, which will enable measurements of the SZ effects with unprecedented precision for thousands of clusters up to high redshifts (e.g., \citealt{Benson2014,Crites2014, Lagache2018,Stacey2018, Mroczkowski2019}). Future space missions, on the other hand, will improve upon the sensitivity and spectral coverage of Planck, pushing the SZ-cluster statistics above the $10^5$-object level (e.g., \citealt{Kogut2016, Delabrouille2018,Melin2018,Smirnov2018}). The potential applications of these data sets in the context of cosmology, the halo--galaxy connection, and the large-scale structure of the Universe (LSS) are numerous and promising (e.g., \citealt{HM2006,Bhattacharya2007,Bhattacharya2008,Zhang2011,Hand2012, Eva2015, Alonso2016, Soergel2018, Basu2019, Mroczkowski2019, Andrina2020}). Our results add 
the study of the secondary dependencies of halo and galaxy bias to this list of future prospects. For a more detailed description of current and future instrumentation and its applications, we refer the reader to \cite{Mroczkowski2019}.

\section{Conclusions}
\label{sec:conclusions}

In this paper, we use the TNG300 hydrodynamical simulation box to explore the observational detectability of {\it{halo spin bias}}, the secondary dependence of halo bias on halo spin at fixed halo mass. In simulations, faster-rotating haloes at $z=0$ are more strongly clustered than slower-rotating haloes of the same mass for halo masses above $\log_{10} ({\rm M_{halo}}/h^{-1} {\rm M_{\odot}}) \simeq 11.5$  (see recent measurements in \citealt{SatoPolito2019, Johnson2019, MonteroDorta2020B, Tucci2020}). For the most massive, cluster-size haloes, the bias ratio between these subpopulations is at least 1.5, as measured from N-body numerical simulations \citep{SatoPolito2019, Johnson2019, Tucci2020}. In this range of masses, the assembly bias signal is expected to be minimal (if it even exists, see discussion in, e.g., \citealt{Mao2018,Chue2018,SatoPolito2019}), which elevates the potential of spin bias as an interesting alternative for probing secondary halo bias. 

Measuring the rotation of haloes from observations is inherently much more challenging than measuring galaxy ages or star formation histories, which could potentially trace the assembly bias signal. However, hydrodynamical simulations show that the total angular momentum of the gas inside the halo (i.e., the intra-halo gas) correlates with that of the halo's DM component (see, e.g., \citealt{MonteroDorta2020B}). Motivated by this result, we measure both the kinetic and thermal Sunyaev Zel'dovich signals (kSZ, and tSZ, respectively) produced by more than 50,000 haloes in the TNG300 box, spanning a wide range of halo masses. The kSZ effect is known to be capable of tracing the angular momentum of the intra-halo gas, whereas the total gas mass of the halo can be recovered from measurements of either the kSZ or the tSZ effects. In this paper, we analyse the extent to which these correlations with the baryonic content of haloes transmit to their virial properties, and take advantage of the {\it{cosmological}} size of TNG300 to measure the dependence of clustering on the integrated SZ signals. The main conclusions of this work can be summarised as follows:

\begin{itemize} 
    \item Both the tSZ and the kSZ integrated signals, represented here by Y$_{\rm tSZ}$ and S$_{\rm kSZ}$, are good tracers of the total virial mass of the halo, ${\rm M_{vir}}$, with the scatter being significantly smaller at the high-mass end. The Y$_{\rm tSZ}$ statistic, in particular, provides a remarkably tight correlation for haloes above $\log_{10} ({\rm M_{vir}}/h^{-1} {\rm M_{\odot}}) > 13$. The Spearman rank-order correlation coefficients for these relations are $\rho = 0.59$ (Y$_{\rm tSZ}$) and $\rho = 0.50$ (S$_{\rm kSZ}$) within the entire halo mass range considered. 
    \item We have evaluated the performance of both S$_{\rm kSZ}$ and the ratio of the integrated kSZ and tSZ signals, L$_{\rm SZ}$, as a proxy for the total halo spin, $\lambda_{\rm halo}$. Both statistics provide a noticeable correlation with $\lambda_{\rm halo}$, although the observed scatter is, as expected, large ($\rho =$ 0.35 for L$_{\rm SZ}$ and 0.33 for S$_{\rm kSZ}$). This reflects the turbulent motion of gas inside the haloes, particularly at the low-mass end.  
    \item We have shown that, in the absence of observational uncertainties, most of the underlying TNG300 halo spin bias signal can be recovered if the SZ-based statistics S$_{\rm kSZ}$ and L$_{\rm SZ}$ are used to split the halo sample at fixed halo mass. The performance of S$_{\rm kSZ}$ in terms of tracing the underlying signal is slightly better than that of L$_{\rm SZ}$, although both measurements are fairly compatible within errors. By using L$_{\rm SZ}$, we can mitigate the potential effect of halo mass on the spin bias signal.
    \item The above statement remains true when the integrated tSZ signal, Y$_{\rm tSZ}$, is used as a proxy for virial mass at the high-mass end. This result emphasises the potential of the tSZ technique in the context of future observational probes of secondary halo bias. 
    \item The above conclusions regarding the clustering properties of kSZ haloes must be corroborated with a larger (simulated) cosmological volume in the future, in order to reduce the statistical errors in relative bias for the most massive haloes. This can be attainable, for instance, with the BAryons and HAloes of MAssive Systems (BAHAMAS) suite of hydrodynamical simulations \citep{McCarthy2017}, which provides a box of 400 $h^{-1}$Mpc on a side, i.e., $\sim$8 times larger in volume than TNG300.
    \item Although the kSZ effect has only been tentatively detected in a few cases, future instrumentation and space missions are expected to deliver large samples of tSZ and kSZ data. Altogether, our results add to the potential applications of the SZ effect as a powerful probe for the LSS and the halo--galaxy connection, once these data sets are made available to the community. 
\end{itemize}

While measuring halo spin for large samples of haloes will be feasible in the future, progress is still needed in terms of understanding the physical origins of the halo spin bias effect from a theoretical standpoint. In \cite{Tucci2020}, we show that the low-mass end is dominated by splashback haloes, which produce the inversion of the signal reported in \cite{SatoPolito2019} and \cite{Johnson2019}. These haloes live in the vicinity of significantly more massive haloes, thus sharing their high large-scale biases. At the high-mass end, on the contrary, the physical mechanisms that determine the intrinsic spin-bias behaviour are still not fully understood. Recent results indicate that the (potentially related) assembly bias trend is connected to the anisotropy of the tidal environment \citep{Paranjape2018,Ramakrishnan2019,Paranjape2020, Ramakrishnan2020}, which in turn seems to align well with seminal theories that attribute the acquisition of angular momentum by haloes to the large-scale tidal field (e.g., \citealt{Peebles1969, Doroshkevich1970, White1984, BarnesEfstathiou1987}).

Finally, investigating the impact that the second-order dependencies of halo clustering have on the galaxy population is relevant in the current era of {\it{precision cosmology}}. Massive cosmological surveys such as the Dark Energy Spectroscopic Instrument (DESI\footnote{\url{https://www.desi.lbl.gov}}), 
Euclid\footnote{\url{https://www.euclid-ec.org}}, the Javalambre Physics of the Accelerated Universe Astrophysical Survey (J-PAS\footnote{\url{http://www.j-pas.org}}), or even the Large Synoptic Spectroscopic Telescope (LSST\footnote{\url{https://www.lsst.org}}) in the future will map the LSS with unprecedented precision, using a variety of cosmological tracers. Optimising the amount of information that can be extracted from these data sets relies on our ability to disentangle the multiple connections between galaxies and the underlying matter density field.

\section*{Acknowledgments}

ADMD and MCA acknowledge Claudia Lagos for useful comments and suggestions at the beginning of this project. ADMD and BT thanks FAPESP for financial support. MCA acknowledges financial support from the Austrian National Science Foundation through FWF stand-alone grant P31154-N27. LRA thanks both FAPESP and CNPq for financial support. This research made use of the {\sc Dirty Astropy} module, developed by Adam Stevens\footnote{ \url{https://github.com/arhstevens/Dirty-AstroPy}}.

\section*{Data availability}
The simulation data underlying this article are publicly
available at the CosmoSim, Skies \& Universes, and TNG websites. The data results arising from this work will be shared on reasonable request to the corresponding authors.

\bibliography{./paper}

\begin{thebibliography}{109}
\expandafter\ifx\csname natexlab\endcsname\relax\def\natexlab#1{#1}\fi

\bibitem[{Adam} et~al.(2017){Adam}, {Bartalucci}, {Pratt} et~al.]{Adam2017}
{Adam} R., {Bartalucci} I., {Pratt} G.~W., et~al., 2017, \aap, 598, A115

\bibitem[{Alonso} et~al.(2016){Alonso}, {Louis}, {Bull} \&
  {Ferreira}]{Alonso2016}
{Alonso} D., {Louis} T., {Bull} P., {Ferreira} P.~G., 2016, \prd, 94, 4, 043522

\bibitem[{Andersson} et~al.(2011){Andersson}, {Benson}, {Ade}
  et~al.]{Andersson2011}
{Andersson} K., {Benson} B.~A., {Ade} P.~A.~R., et~al., 2011, \apj, 738, 1, 48

\bibitem[{Angulo} et~al.(2008){Angulo}, {Baugh} \& {Lacey}]{Angulo2008}
{Angulo} R.~E., {Baugh} C.~M., {Lacey} C.~G., 2008, \mnras, 387, 921

\bibitem[{Arnaud} et~al.(2010){Arnaud}, {Pratt}, {Piffaretti}, {B{\"o}hringer},
  {Croston} \& {Pointecouteau}]{Arnaud2010}
{Arnaud} M., {Pratt} G.~W., {Piffaretti} R., {B{\"o}hringer} H., {Croston}
  J.~H., {Pointecouteau} E., 2010, \aap, 517, A92

\bibitem[{Baldi} et~al.(2018){Baldi}, {De Petris}, {Sembolini}, {Yepes}, {Cui}
  \& {Lamagna}]{Baldi2018}
{Baldi} A.~S., {De Petris} M., {Sembolini} F., {Yepes} G., {Cui} W., {Lamagna}
  L., 2018, \mnras, 479, 3, 4028

\bibitem[{Barnes} \& {Efstathiou}(1987)]{BarnesEfstathiou1987}
{Barnes} J., {Efstathiou} G., 1987, \apj, 319, 575

\bibitem[{Basu} et~al.(2019){Basu}, {Erler}, {Chluba} et~al.]{Basu2019}
{Basu} K., {Erler} J., {Chluba} J., et~al., 2019, \baas, 51, 3, 302

\bibitem[{Battaglia} et~al.(2012){Battaglia}, {Bond}, {Pfrommer} \&
  {Sievers}]{Battaglia2012}
{Battaglia} N., {Bond} J.~R., {Pfrommer} C., {Sievers} J.~L., 2012, \apj, 758,
  2, 74

\bibitem[{Behroozi} et~al.(2013){Behroozi}, {Wechsler} \&
  {Conroy}]{Behroozi2013}
{Behroozi} P.~S., {Wechsler} R.~H., {Conroy} C., 2013, \apj, 770, 57

\bibitem[{Benson} et~al.(2014){Benson}, {Ade}, {Ahmed} et~al.]{Benson2014}
{Benson} B.~A., {Ade} P.~A.~R., {Ahmed} Z., et~al., 2014, in { Millimeter,
  Submillimeter, and Far-Infrared Detectors and Instrumentation for Astronomy
  VII\/}, vol. 9153 of { Society of Photo-Optical Instrumentation Engineers
  (SPIE) Conference Series\/},  91531P

\bibitem[{Bhattacharya} \& {Kosowsky}(2007)]{Bhattacharya2007}
{Bhattacharya} S., {Kosowsky} A., 2007, \apjl, 659, 2, L83

\bibitem[{Bhattacharya} \& {Kosowsky}(2008)]{Bhattacharya2008}
{Bhattacharya} S., {Kosowsky} A., 2008, \prd, 77, 8, 083004

\bibitem[{Birkinshaw} et~al.(1984){Birkinshaw}, {Gull} \&
  {Hardebeck}]{Birkinshaw1984}
{Birkinshaw} M., {Gull} S.~F., {Hardebeck} H., 1984, \nat, 309, 34

\bibitem[{Bleem} et~al.(2015){Bleem}, {Stalder}, {de Haan} et~al.]{Bleem2015}
{Bleem} L.~E., {Stalder} B., {de Haan} T., et~al., 2015, \apjs, 216, 2, 27

\bibitem[{Bocquet} et~al.(2019){Bocquet}, {Dietrich}, {Schrabback}
  et~al.]{Bocquet2019}
{Bocquet} S., {Dietrich} J.~P., {Schrabback} T., et~al., 2019, \apj, 878, 1, 55

\bibitem[{Bolliet} et~al.(2020){Bolliet}, {Brinckmann}, {Chluba} \&
  {Lesgourgues}]{Bolliet2020}
{Bolliet} B., {Brinckmann} T., {Chluba} J., {Lesgourgues} J., 2020, \mnras,
  497, 2, 1332

\bibitem[{Bolliet} et~al.(2018){Bolliet}, {Comis}, {Komatsu} \&
  {Mac{\'\i}as-P{\'e}rez}]{Bolliet2018}
{Bolliet} B., {Comis} B., {Komatsu} E., {Mac{\'\i}as-P{\'e}rez} J.~F., 2018,
  \mnras, 477, 4, 4957

\bibitem[{Bullock} et~al.(2001){Bullock}, {Dekel}, {Kolatt}
  et~al.]{Bullock2001}
{Bullock} J.~S., {Dekel} A., {Kolatt} T.~S., et~al., 2001, \apj, 555, 1, 240

\bibitem[{Chluba} \& {Mannheim}(2002)]{Chluba2002}
{Chluba} J., {Mannheim} K., 2002, \aap, 396, 419

\bibitem[{Chue} et~al.(2018){Chue}, {Dalal} \& {White}]{Chue2018}
{Chue} C. Y.~R., {Dalal} N., {White} M., 2018, \jcap, 2018, 10, 012

\bibitem[{Cooray} \& {Chen}(2002)]{Cooray2002}
{Cooray} A., {Chen} X., 2002, \apj, 573, 1, 43

\bibitem[{Crites} et~al.(2014){Crites}, {Bock}, {Bradford} et~al.]{Crites2014}
{Crites} A.~T., {Bock} J.~J., {Bradford} C.~M., et~al., 2014, in { Millimeter,
  Submillimeter, and Far-Infrared Detectors and Instrumentation for Astronomy
  VII\/}, vol. 9153 of { Society of Photo-Optical Instrumentation Engineers
  (SPIE) Conference Series\/},  91531W

\bibitem[{Davis} et~al.(1985){Davis}, {Efstathiou}, {Frenk} \&
  {White}]{Davis1985}
{Davis} M., {Efstathiou} G., {Frenk} C.~S., {White} S.~D.~M., 1985, \apj, 292,
  371

\bibitem[{Delabrouille} et~al.(2018){Delabrouille}, {de Bernardis}, {Bouchet}
  et~al.]{Delabrouille2018}
{Delabrouille} J., {de Bernardis} P., {Bouchet} F.~R., et~al., 2018, \jcap,
  2018, 4, 014

\bibitem[{Dolag} et~al.(2009){Dolag}, {Borgani}, {Murante} \&
  {Springel}]{Dolag2009}
{Dolag} K., {Borgani} S., {Murante} G., {Springel} V., 2009, \mnras, 399, 2,
  497

\bibitem[{Doroshkevich}(1970)]{Doroshkevich1970}
{Doroshkevich} A.~G., 1970, Astrophysics, 6, 4, 320

\bibitem[{Dupke} \& {Bregman}(2002)]{Dupke2002}
{Dupke} R.~A., {Bregman} J.~N., 2002, \apj, 575, 2, 634

\bibitem[{Fixsen}(2009)]{Fixsen2009}
{Fixsen} D.~J., 2009, \apj, 707, 2, 916

\bibitem[{Gao} et~al.(2005){Gao}, {Springel} \& {White}]{gao2005}
{Gao} L., {Springel} V., {White} S.~D.~M., 2005, \mnras, 363, L66

\bibitem[{Gao} \& {White}(2007)]{Gao2007}
{Gao} L., {White} S.~D.~M., 2007, \mnras, 377, L5

\bibitem[{Han} et~al.(2018){Han}, {Li}, {Jing}, {Nishimichi}, {Wang} \&
  {Jiang}]{han2018}
{Han} J., {Li} Y., {Jing} Y., {Nishimichi} T., {Wang} W., {Jiang} C., 2018,
  ArXiv e-prints

\bibitem[{Hand} et~al.(2012){Hand}, {Addison}, {Aubourg} et~al.]{Hand2012}
{Hand} N., {Addison} G.~E., {Aubourg} E., et~al., 2012, \prl, 109, 4, 041101

\bibitem[{Hasselfield} et~al.(2013){Hasselfield}, {Hilton}, {Marriage}
  et~al.]{Hasselfield2013}
{Hasselfield} M., {Hilton} M., {Marriage} T.~A., et~al., 2013, \jcap, 2013, 7,
  008

\bibitem[{Hearin} \& {Watson}(2013)]{Hearin2013}
{Hearin} A.~P., {Watson} D.~F., 2013, \mnras, 435, 1313

\bibitem[{Hearin} et~al.(2014){Hearin}, {Watson}, {Becker}, {Reyes}, {Berlind}
  \& {Zentner}]{Hearin2014}
{Hearin} A.~P., {Watson} D.~F., {Becker} M.~R., {Reyes} R., {Berlind} A.~A.,
  {Zentner} A.~R., 2014, \mnras, 444, 729

\bibitem[{Hearin} et~al.(2016){Hearin}, {Zentner}, {van den Bosch}, {Campbell}
  \& {Tollerud}]{Hearin2016}
{Hearin} A.~P., {Zentner} A.~R., {van den Bosch} F.~C., {Campbell} D.,
  {Tollerud} E., 2016, \mnras, 460, 2552

\bibitem[{Hern{\'a}ndez-Monteagudo} et~al.(2006){Hern{\'a}ndez-Monteagudo},
  {Verde}, {Jimenez} \& {Spergel}]{HM2006}
{Hern{\'a}ndez-Monteagudo} C., {Verde} L., {Jimenez} R., {Spergel} D.~N., 2006,
  \apj, 643, 2, 598

\bibitem[{Hurier} \& {Lacasa}(2017)]{Hurier2017}
{Hurier} G., {Lacasa} F., 2017, \aap, 604, A71

\bibitem[{Johnson} et~al.(2019){Johnson}, {Maller}, {Berlind}, {Sinha} \&
  {Holley-Bockelmann}]{Johnson2019}
{Johnson} J.~W., {Maller} A.~H., {Berlind} A.~A., {Sinha} M.,
  {Holley-Bockelmann} J.~K., 2019, \mnras, 486, 1, 1156

\bibitem[{Klypin} et~al.(2016){Klypin}, {Yepes}, {Gottl{\"o}ber}, {Prada} \&
  {He{\ss}}]{multidark2016}
{Klypin} A., {Yepes} G., {Gottl{\"o}ber} S., {Prada} F., {He{\ss}} S., 2016,
  \mnras, 457, 4340

\bibitem[{Kogut} et~al.(2016){Kogut}, {Chluba}, {Fixsen}, {Meyer} \&
  {Spergel}]{Kogut2016}
{Kogut} A., {Chluba} J., {Fixsen} D.~J., {Meyer} S., {Spergel} D., 2016, in {
  Space Telescopes and Instrumentation 2016: Optical, Infrared, and Millimeter
  Wave\/}, vol. 9904 of { Society of Photo-Optical Instrumentation Engineers
  (SPIE) Conference Series\/},  99040W

\bibitem[{Krause} et~al.(2012){Krause}, {Pierpaoli}, {Dolag} \&
  {Borgani}]{Krause2012}
{Krause} E., {Pierpaoli} E., {Dolag} K., {Borgani} S., 2012, \mnras, 419, 2,
  1766

\bibitem[{Lacerna} \& {Padilla}(2012)]{Lacerna2012}
{Lacerna} I., {Padilla} N., 2012, \mnras, 426, 1, L26

\bibitem[{Lagache}(2018)]{Lagache2018}
{Lagache} G., 2018, in { Peering towards Cosmic Dawn\/}, edited by
  V.~{Jeli{\'c}}, T.~{van der Hulst}, vol. 333 of { IAU Symposium\/},  228--233

\bibitem[{Lazeyras} et~al.(2017){Lazeyras}, {Musso} \& {Schmidt}]{Lazeyras2017}
{Lazeyras} T., {Musso} M., {Schmidt} F., 2017, \jcap, 2017, 3, 059

\bibitem[{Li} et~al.(2008){Li}, {Mo} \& {Gao}]{li2008}
{Li} Y., {Mo} H.~J., {Gao} L., 2008, \mnras, 389, 1419

\bibitem[{Li} et~al.(2018){Li}, {Ma}, {Remazeilles} \& {Moodley}]{Li_kSZ2018}
{Li} Y.-C., {Ma} Y.-Z., {Remazeilles} M., {Moodley} K., 2018, \prd, 97, 2,
  023514

\bibitem[{Lim} et~al.(2020){Lim}, {Barnes}, {Vogelsberger} et~al.]{Lim2020}
{Lim} S.~H., {Barnes} D., {Vogelsberger} M., et~al., 2020, arXiv e-prints,
  arXiv:2007.11583

\bibitem[{Lin} et~al.(2016){Lin}, {Mandelbaum}, {Huang} et~al.]{Lin2016}
{Lin} Y.-T., {Mandelbaum} R., {Huang} Y.-H., et~al., 2016, \apj, 819, 119

\bibitem[{Mansfield} \& {Kravtsov}(2020)]{Mansfield2020}
{Mansfield} P., {Kravtsov} A.~V., 2020, \mnras, 493, 4, 4763

\bibitem[{Mao} et~al.(2018){Mao}, {Zentner} \& {Wechsler}]{Mao2018}
{Mao} Y.-Y., {Zentner} A.~R., {Wechsler} R.~H., 2018, \mnras, 474, 4, 5143

\bibitem[{Marinacci} et~al.(2017){Marinacci}, {Grand}, {Pakmor}
  et~al.]{Marinacci2017}
{Marinacci} F., {Grand} R. J.~J., {Pakmor} R., et~al., 2017, \mnras, 466, 4,
  3859

\bibitem[{Marriage} et~al.(2011){Marriage}, {Acquaviva}, {Ade}
  et~al.]{Marriage2011}
{Marriage} T.~A., {Acquaviva} V., {Ade} P. A.~R., et~al., 2011, \apj, 737, 2,
  61

\bibitem[{Mather} et~al.(1999){Mather}, {Fixsen}, {Shafer}, {Mosier} \&
  {Wilkinson}]{Mather1999}
{Mather} J.~C., {Fixsen} D.~J., {Shafer} R.~A., {Mosier} C., {Wilkinson} D.~T.,
  1999, \apj, 512, 2, 511

\bibitem[{McCarthy} et~al.(2017){McCarthy}, {Schaye}, {Bird} \& {Le
  Brun}]{McCarthy2017}
{McCarthy} I.~G., {Schaye} J., {Bird} S., {Le Brun} A. M.~C., 2017, \mnras,
  465, 3, 2936

\bibitem[{Melin} et~al.(2018){Melin}, {Bonaldi}, {Remazeilles}
  et~al.]{Melin2018}
{Melin} J.~B., {Bonaldi} A., {Remazeilles} M., et~al., 2018, \jcap, 2018, 4,
  019

\bibitem[{Miyatake} et~al.(2016){Miyatake}, {More}, {Takada}
  et~al.]{Miyatake2016}
{Miyatake} H., {More} S., {Takada} M., et~al., 2016, Physical Review Letters,
  116, 4, 041301

\bibitem[{Montero-Dorta} et~al.(2020){Montero-Dorta}, {Artale}, {Abramo}
  et~al.]{MonteroDorta2020B}
{Montero-Dorta} A.~D., {Artale} M.~C., {Abramo} L.~R., et~al., 2020, \mnras,
  496, 2, 1182

\bibitem[{Montero-Dorta} et~al.(2017){Montero-Dorta}, {P{\'e}rez}, {Prada}
  et~al.]{MonteroDorta2017B}
{Montero-Dorta} A.~D., {P{\'e}rez} E., {Prada} F., et~al., 2017, \apjl, 848, L2

\bibitem[{Mroczkowski} et~al.(2012){Mroczkowski}, {Dicker}, {Sayers}
  et~al.]{Mroczkowski2012}
{Mroczkowski} T., {Dicker} S., {Sayers} J., et~al., 2012, \apj, 761, 1, 47

\bibitem[{Mroczkowski} et~al.(2019){Mroczkowski}, {Nagai}, {Basu}
  et~al.]{Mroczkowski2019}
{Mroczkowski} T., {Nagai} D., {Basu} K., et~al., 2019, \ssr, 215, 1, 17

\bibitem[{Mueller} et~al.(2015){Mueller}, {de Bernardis}, {Bean} \&
  {Niemack}]{Eva2015}
{Mueller} E.-M., {de Bernardis} F., {Bean} R., {Niemack} M.~D., 2015, \apj,
  808, 1, 47

\bibitem[{Nicola} et~al.(2020){Nicola}, {Dunkley} \& {Spergel}]{Andrina2020}
{Nicola} A., {Dunkley} J., {Spergel} D.~N., 2020, arXiv e-prints,
  arXiv:2006.00008

\bibitem[{Niemiec} et~al.(2018){Niemiec}, {Jullo}, {Montero-Dorta}
  et~al.]{Niemiec2018}
{Niemiec} A., {Jullo} E., {Montero-Dorta} A.~D., et~al., 2018, \mnras

\bibitem[{Obuljen} et~al.(2020){Obuljen}, {Percival} \& {Dalal}]{Obuljen2020}
{Obuljen} A., {Percival} W.~J., {Dalal} N., 2020, arXiv e-prints,
  arXiv:2004.07240

\bibitem[{Paranjape}(2020)]{Paranjape2020}
{Paranjape} A., 2020, arXiv e-prints,  arXiv:2006.13954

\bibitem[{Paranjape} et~al.(2018){Paranjape}, {Hahn} \& {Sheth}]{Paranjape2018}
{Paranjape} A., {Hahn} O., {Sheth} R.~K., 2018, \mnras, 476, 3, 3631

\bibitem[{Peebles}(1969)]{Peebles1969}
{Peebles} P.~J.~E., 1969, \apj, 155, 393

\bibitem[{Pillepich} et~al.(2018){Pillepich}, {Springel}, {Nelson}
  et~al.]{Pillepich2018}
{Pillepich} A., {Springel} V., {Nelson} D., et~al., 2018, \mnras, 473, 3, 4077

\bibitem[{Planck Collaboration} et~al.(2014){Planck Collaboration}, {Ade},
  {Aghanim} et~al.]{Planck2013}
{Planck Collaboration}, {Ade} P.~A.~R., {Aghanim} N., et~al., 2014, \aap, 571,
  A16

\bibitem[{Planck Collaboration} et~al.(2016{\natexlab{a}}){Planck
  Collaboration}, {Ade}, {Aghanim} et~al.]{Planck2016}
{Planck Collaboration}, {Ade} P.~A.~R., {Aghanim} N., et~al.,
  2016{\natexlab{a}}, \aap, 594, A13

\bibitem[{Planck Collaboration} et~al.(2016{\natexlab{b}}){Planck
  Collaboration}, {Ade}, {Aghanim} et~al.]{Planck_kSZ2016}
{Planck Collaboration}, {Ade} P.~A.~R., {Aghanim} N., et~al.,
  2016{\natexlab{b}}, \aap, 586, A140

\bibitem[{Ramakrishnan} \& {Paranjape}(2020)]{Ramakrishnan2020}
{Ramakrishnan} S., {Paranjape} A., 2020, arXiv e-prints,  arXiv:2007.03711

\bibitem[{Ramakrishnan} et~al.(2019){Ramakrishnan}, {Paranjape}, {Hahn} \&
  {Sheth}]{Ramakrishnan2019}
{Ramakrishnan} S., {Paranjape} A., {Hahn} O., {Sheth} R.~K., 2019, \mnras, 489,
  3, 2977

\bibitem[{Rephaeli}(1995)]{Rephaeli1995}
{Rephaeli} Y., 1995, \araa, 33, 541

\bibitem[{Salcedo} et~al.(2018){Salcedo}, {Maller}, {Berlind}
  et~al.]{salcedo2018}
{Salcedo} A.~N., {Maller} A.~H., {Berlind} A.~A., et~al., 2018, \mnras, 475,
  4411

\bibitem[{Sato-Polito} et~al.(2019){Sato-Polito}, {Montero-Dorta}, {Abramo},
  {Prada} \& {Klypin}]{SatoPolito2019}
{Sato-Polito} G., {Montero-Dorta} A.~D., {Abramo} L.~R., {Prada} F., {Klypin}
  A., 2019, \mnras, 487, 2, 1570

\bibitem[{Sayers} et~al.(2019){Sayers}, {Monta{\~n}a}, {Mroczkowski}
  et~al.]{Sayers2019}
{Sayers} J., {Monta{\~n}a} A., {Mroczkowski} T., et~al., 2019, \apj, 880, 1, 45

\bibitem[{Sayers} et~al.(2013){Sayers}, {Mroczkowski}, {Zemcov}
  et~al.]{Sayers2013}
{Sayers} J., {Mroczkowski} T., {Zemcov} M., et~al., 2013, \apj, 778, 1, 52

\bibitem[{Schaan} et~al.(2016){Schaan}, {Ferraro}, {Vargas-Maga{\~n}a}
  et~al.]{Schaan2016}
{Schaan} E., {Ferraro} S., {Vargas-Maga{\~n}a} M., et~al., 2016, \prd, 93, 8,
  082002

\bibitem[{Sembolini} et~al.(2013){Sembolini}, {Yepes}, {De Petris},
  {Gottl{\"o}ber}, {Lamagna} \& {Comis}]{Sembolini2013}
{Sembolini} F., {Yepes} G., {De Petris} M., {Gottl{\"o}ber} S., {Lamagna} L.,
  {Comis} B., 2013, \mnras, 429, 1, 323

\bibitem[{Sheth} \& {Tormen}(2004)]{Sheth2004}
{Sheth} R.~K., {Tormen} G., 2004, \mnras, 350, 4, 1385

\bibitem[{Sif{\'o}n} et~al.(2013){Sif{\'o}n}, {Menanteau}, {Hasselfield}
  et~al.]{Sifon2013}
{Sif{\'o}n} C., {Menanteau} F., {Hasselfield} M., et~al., 2013, \apj, 772, 1,
  25

\bibitem[{Smirnov} et~al.(2018){Smirnov}, {Pilipenko}, {Golubev}
  et~al.]{Smirnov2018}
{Smirnov} A., {Pilipenko} S., {Golubev} E., et~al., 2018, in { 42nd COSPAR
  Scientific Assembly\/}, vol.~42,  E4.2--12--18

\bibitem[{Soergel} et~al.(2018){Soergel}, {Saro}, {Giannantonio}, {Efstathiou}
  \& {Dolag}]{Soergel2018}
{Soergel} B., {Saro} A., {Giannantonio} T., {Efstathiou} G., {Dolag} K., 2018,
  \mnras, 478, 4, 5320

\bibitem[{Springel}(2010)]{Springel2010}
{Springel} V., 2010, \mnras, 401, 2, 791

\bibitem[{Springel} \& {Hernquist}(2003)]{springel03}
{Springel} V., {Hernquist} L., 2003, \mnras, 339, 2, 289

\bibitem[{Springel} et~al.(2018){Springel}, {Pakmor}, {Pillepich}
  et~al.]{Springel2018}
{Springel} V., {Pakmor} R., {Pillepich} A., et~al., 2018, \mnras, 475, 1, 676

\bibitem[{Springel} et~al.(2001){Springel}, {White}, {Tormen} \&
  {Kauffmann}]{Springel2001}
{Springel} V., {White} S. D.~M., {Tormen} G., {Kauffmann} G., 2001, \mnras,
  328, 3, 726

\bibitem[{Stacey} et~al.(2018){Stacey}, {Aravena}, {Basu} et~al.]{Stacey2018}
{Stacey} G.~J., {Aravena} M., {Basu} K., et~al., 2018, in { Ground-based and
  Airborne Telescopes VII\/}, vol. 10700 of { Society of Photo-Optical
  Instrumentation Engineers (SPIE) Conference Series\/},  107001M

\bibitem[{Stevens} et~al.(2019){Stevens}, {Diemer}, {Lagos}
  et~al.]{Stevens2019}
{Stevens} A. R.~H., {Diemer} B., {Lagos} C. d.~P., et~al., 2019, \mnras, 483,
  4, 5334

\bibitem[{Sunayama} et~al.(2016){Sunayama}, {Hearin}, {Padmanabhan} \&
  {Leauthaud}]{Sunayama2016}
{Sunayama} T., {Hearin} A.~P., {Padmanabhan} N., {Leauthaud} A., 2016, \mnras,
  458, 1510

\bibitem[{Sunayama} \& {More}(2019)]{Sunayama2019}
{Sunayama} T., {More} S., 2019, \mnras, 490, 4, 4945

\bibitem[{Sunyaev} \& {Zel'dovich}(1980{\natexlab{a}})]{SZ1980A}
{Sunyaev} R.~A., {Zel'dovich} I.~B., 1980{\natexlab{a}}, \araa, 18, 537

\bibitem[{Sunyaev} \& {Zel'dovich}(1980{\natexlab{b}})]{SZ1980B}
{Sunyaev} R.~A., {Zel'dovich} I.~B., 1980{\natexlab{b}}, \mnras, 190, 413

\bibitem[{Sunyaev} \& {Zel'dovich}(1970)]{SZ1970}
{Sunyaev} R.~A., {Zel'dovich} Y.~B., 1970, \apss, 7, 1, 20

\bibitem[{Tucci} et~al.(2020){Tucci}, {Montero-Dorta}, {Abramo}, {Sato-Polito}
  \& {Artale}]{Tucci2020}
{Tucci} B., {Montero-Dorta} A.~D., {Abramo} L.~R., {Sato-Polito} G., {Artale}
  M.~C., 2020, arXiv e-prints,  arXiv:2007.10366

\bibitem[{Vogelsberger} et~al.(2014{\natexlab{a}}){Vogelsberger}, {Genel},
  {Springel} et~al.]{Vogelsberger2014a}
{Vogelsberger} M., {Genel} S., {Springel} V., et~al., 2014{\natexlab{a}},
  \mnras, 444, 2, 1518

\bibitem[{Vogelsberger} et~al.(2014{\natexlab{b}}){Vogelsberger}, {Genel},
  {Springel} et~al.]{Vogelsberger2014b}
{Vogelsberger} M., {Genel} S., {Springel} V., et~al., 2014{\natexlab{b}}, \nat,
  509, 7499, 177

\bibitem[{Walsh} \& {Tinker}(2019)]{Walsh2019}
{Walsh} K., {Tinker} J., 2019, \mnras, 488, 1, 470

\bibitem[{Wechsler} \& {Tinker}(2018)]{Wechsler2018}
{Wechsler} R.~H., {Tinker} J.~L., 2018, \araa, 56, 435

\bibitem[{Wechsler} et~al.(2006){Wechsler}, {Zentner}, {Bullock}, {Kravtsov} \&
  {Allgood}]{wechsler2006}
{Wechsler} R.~H., {Zentner} A.~R., {Bullock} J.~S., {Kravtsov} A.~V., {Allgood}
  B., 2006, \apj, 652, 71

\bibitem[{White}(1984)]{White1984}
{White} S.~D.~M., 1984, \apj, 286, 38

\bibitem[{Yu} et~al.(2015){Yu}, {Nelson} \& {Nagai}]{Yu2015}
{Yu} L., {Nelson} K., {Nagai} D., 2015, \apj, 807, 1, 12

\bibitem[{Zentner} et~al.(2016){Zentner}, {Hearin}, {van den Bosch}, {Lange} \&
  {Villarreal}]{Zentner2016}
{Zentner} A.~R., {Hearin} A., {van den Bosch} F.~C., {Lange} J.~U.,
  {Villarreal} A., 2016, ArXiv e-prints

\bibitem[{Zentner} et~al.(2019){Zentner}, {Hearin}, {van den Bosch}, {Lange} \&
  {Villarreal}]{Zentner2019}
{Zentner} A.~R., {Hearin} A., {van den Bosch} F.~C., {Lange} J.~U.,
  {Villarreal} A., 2019, \mnras, 485, 1, 1196

\bibitem[{Zhang} \& {Stebbins}(2011)]{Zhang2011}
{Zhang} P., {Stebbins} A., 2011, \prl, 107, 4, 041301

\bibitem[{Zu} et~al.(2016){Zu}, {Mandelbaum}, {Simet}, {Rozo} \&
  {Rykoff}]{Zu2016}
{Zu} Y., {Mandelbaum} R., {Simet} M., {Rozo} E., {Rykoff} E.~S., 2016, ArXiv
  e-prints

\end{thebibliography}

\appendix

\section{The SZ effects from TNG100}
\label{sec:appendix_TNG100}

\begin{figure}
%\centering
    \includegraphics[width=\columnwidth]{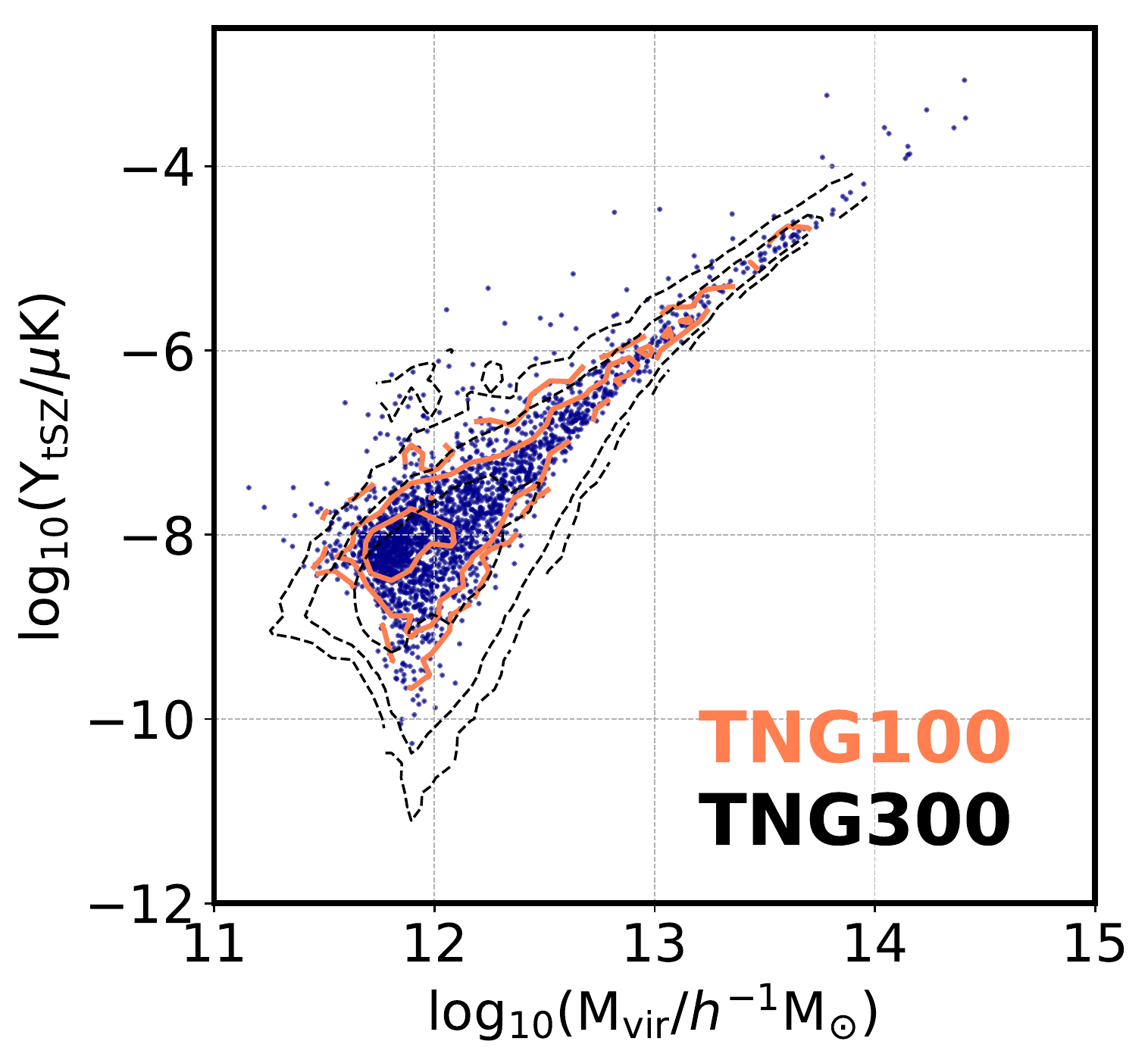}
    \caption{The relation between the logarithm of the integrated tSZ signal, Y$_{\rm tSZ}$, and the logarithm of the total virial mass of the halo, M$_{\rm vir}$, as measured from TNG100 (blue dots, orange contours) and TNG300 (black contours).  }
    \label{fig:appendix1}
\end{figure}

\begin{figure*}
%\centering
    \includegraphics[width=\textwidth]{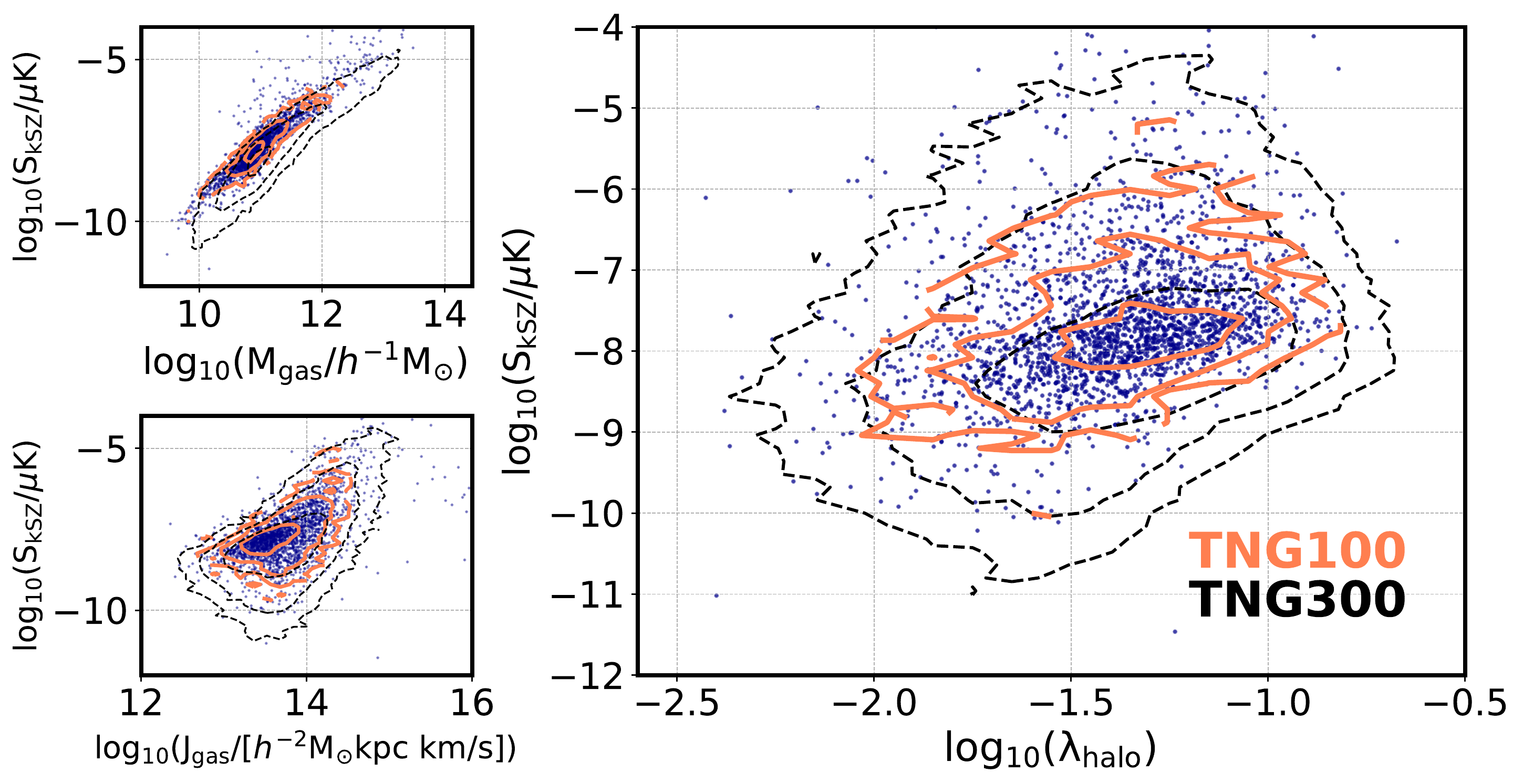}
    \caption{A comparison between the kSZ measurements performed in TNG100 (blue dots, orange contours) and TNG300 (black contours).  The correlations between the logarithm of the integrated kSZ signal, S$_{\rm kSZ}$, and the logarithm of the total gas mass, M$_{\rm gas}$, the total angular momentum of the gas, J$_{\rm gas}$, and the total halo spin, $\lambda_{\rm halo}$, are shown for both boxes.}
    \label{fig:appendix2}
\end{figure*}

In order to test the robustness of the computation of the tSZ and kSZ signals, we have also applied the method to the TNG100 box. As explained in Section~\ref{sec:sims},  TNG100 is significantly smaller than TNG300, with a side length of only 75 $h^{-1}$Mpc, as compared to 205 $h^{-1}$Mpc for TNG300 (a factor $\sim$20 in volume). It is, therefore, not suitable for clustering studies, but it does provide much higher resolution. The masses of individual DM particles and initial gas cells are $5.1\times10^6$ $h^{-1}$M$_{\odot}$ and $9.4 \times 10^5$ $h^{-1}$M$_{\odot}$, respectively, which are roughly an order of magnitude smaller than those of TNG300. Higher mass resolution implies a more realistic description of the dynamics and composition of the baryonic component of the haloes, which is a crucial element in the kSZ and tSZ computations.

We start by comparing, in Figure~\ref{fig:appendix1}, the relation between the integrated tSZ signal, Y$_{\rm tSZ}$, and the total virial mass M$_{\rm vir}$, measured from TNG300 (51,530 haloes) and from TNG100 (2,482 haloes).
Although these are independent simulations, we can compare their distributions and correlations. Figure~\ref{fig:appendix1} displays an excellent agreement between the measurements, despite the difference in resolution. The only discrepancies observed are due to the different halo mass ranges mapped by the two boxes (due to the different volumes). The scatter in the relations is also smaller in TNG100, likely due to the lower number of haloes in this box.

Figure~\ref{fig:appendix2}, on the other hand, demonstrates that the values of the integrated kSZ signal, S$_{\rm kSZ}$, and the correlations with the total mass of the gas, M$_{\rm gas}$, the total angular momentum of the gas, J$_{\rm gas}$, and the total halo spin, $\lambda_{\rm halo}$, are also totally consistent between TNG100 and TNG300. The only differences are, again, the range and scatter of the distributions. Similar agreement is found when the ratio of statistics, L$_{\rm SZ}$, is compared. 

The level of agreement described above is not surprising, since we have effectively reduced the resolution of our haloes by performing the integration in Equation~\ref{eq:kSZ} assuming cells that exceeds the typical size of the gas cells (this reduces drastically the computation time and simplifies the calculation). Figures~\ref{fig:appendix1} and ~\ref{fig:appendix2} make us confident about the robustness of the procedure.

\label{lastpage}

\end{document}